\documentclass[manuscript,screen]{acmart}
\usepackage{caption}
\usepackage{subfigure}
\usepackage[T1]{fontenc}

\usepackage{amsmath,amsfonts}
\usepackage{algorithmic}
\usepackage{graphicx}
\usepackage{textcomp}
\usepackage[dvipsnames]{xcolor}
\usepackage{color, colortbl}
\definecolor{Gray}{gray}{0.9}
\usepackage{amsmath}
\usepackage{pifont}

\usepackage{makecell}
\usepackage{wrapfig}
\usepackage{picinpar}

\usepackage[utf8]{inputenc} %
\usepackage{hyperref}       %
\usepackage{url}            %
\usepackage{booktabs}       %
\usepackage{amsfonts}       %
\usepackage{nicefrac}       %
\usepackage{microtype}      %

\usepackage{algorithmic}
\usepackage{graphicx}
\usepackage{textcomp}
\usepackage{xcolor}
\usepackage{multirow}
\usepackage{algorithm}
\usepackage{bm}
\usepackage{subfigure}
\usepackage{caption}
\usepackage{subcaption}
\usepackage{makecell}

\usepackage{CJKutf8}
\usepackage[T1]{fontenc}
\AtBeginDocument{%
  }

\setcopyright{acmlicensed}
\copyrightyear{2026}
\acmYear{2026}
\acmDOI{10.1145/3816732}

\acmJournal{TOIS}
\acmVolume{44}
\acmNumber{5}
\acmArticle{121}
\acmMonth{7}

\begin{document}

\title{Cassette: Case-to-Case Structural Distillation for Efficient Legal Case Retrieval}

\author{Yanran Tang}
\orcid{0009-0007-9485-1066}
\email{yanran.tang@uq.edu.au}
\author{Ruihong Qiu}
\orcid{0000-0001-8349-6475}
\email{r.qiu@uq.edu.au}
\author{Hongzhi Yin}
\orcid{0000-0003-1395-261X}
\email{h.yin1@uq.edu.au}
\author{Xue Li}
\orcid{0000-0002-4515-6792}
\email{xueli@eecs.uq.edu.au}
\author{Zi Huang}
\orcid{0000-0002-9738-4949}
\email{helen.huang@uq.edu.au}
\affiliation{
  \institution{School of EECS, The University of Queensland}
  \country{Brisbane, Australia}
}

\renewcommand{\shortauthors}{Tang et al.}

\begin{abstract}
Legal case retrieval (LCR) is an essential tool for not only assisting legal practitioners to efficiently retrieve precedents, but also enabling ordinary individuals to find valuable legal case information without relying on expensive professional legal service. Our previous work CaseLink~\cite{caselink} demonstrated the effectiveness of using case to case graph structures to improve retrieval accuracy. However, its high computational cost during inference on large scale legal databases limits its practical use in real world settings. The main inefficiency comes from constructing test time graphs and computing pairwise term frequency similarities of cases. This process has $O(n^2)$ complexity for $n$ legal cases, making the runtime prohibitive as the number of candidates grows.
For example, the retrieval time for one query on a database (COLIEE2022~\cite{COLIEE2022}) with \textbf{\textit{1,563 candidate cases}} is more than \textbf{\textit{500 milliseconds}}, while the runtime would increase drastically to more than \textbf{\textit{3,500 seconds}} for a database (LeCaRDv2~\cite{lecardv2}) with \textbf{\textit{55,192 candidate cases}}. To further enhance the retrieval performance while achieving a significant speed-up, in this extension paper, Cassette framework is proposed with a distillation strategy involving ranking objective and eigen-matching objective for an effective transfer of knowledge from a powerful and well-trained heavy teacher retriever to a lightweight and efficient hybrid student dual encoder. Specifically, the student query encoder is implemented as a multilayer perceptron model designed for fast online processing, whereas the student candidate encoder adopts a GNN architecture, suitable for an offline manner within the case database. Extensive experiments are conducted on three benchmark datasets and the results verify the effectiveness of the ranking distillation while achieving high efficiency. The code has been released on \url{https://github.com/yanran-tang/Cassette}.
\end{abstract}




\maketitle

\section{Introduction}
Legal case retrieval (LCR) is a vital tool that facilitates the identification of relevant cases from a large legal database based on a given query case, providing substantial benefits to both legal professionals and the general public. In legal terminology, these relevant cases are known as precedents, which play a pivotal role in most legal systems worldwide by promoting consistency and fairness in judicial decision-making, as well as strengthening the arguments presented by legal practitioners~\cite{precedent}.

Existing LCR methods can be broadly categorised into two main types: statistical models and neural network models. (1) Statistical retrieval methods, also referred to as lexical models, such as TF-IDF~\cite{TF-IDF}, BM25~\cite{BM25}, and LMIR~\cite{LMIR}, rely on document term frequency to compute similarity scores. (2) Neural network-based methods, by contrast,  leverage deep learning architectures to encode the semantic representations of legal cases and assess relevance, typically within a two-tower framework~\cite{Law2Vec,Lawformer,MTFT-BERT,MVCL,BERT-PLI,LEGAL-BERT,SAILER,JOTR,DoSSIER,RPRS,NOWJ,UA@COLIEE2022,promptcase,casegnn}. More recently, CaseLink~\cite{caselink} has achieved state-of-the-art (SOTA) performance explicitly modelling inter-case relationships using graph neural networks (GNNs), thereby enabling effective message passing across a constructed case graph. 
\begin{figure}[!t]
\centering
    \subfigure[COLIEE2022]{
    \includegraphics[width=0.47\linewidth]{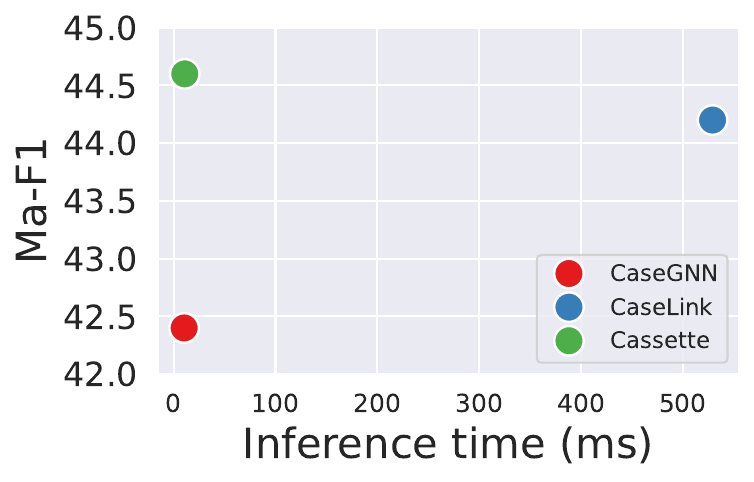}
    \label{fig:fig1_coliee}
    }
    \subfigure[LeCaRDv2]{
    \includegraphics[width=0.47\linewidth]{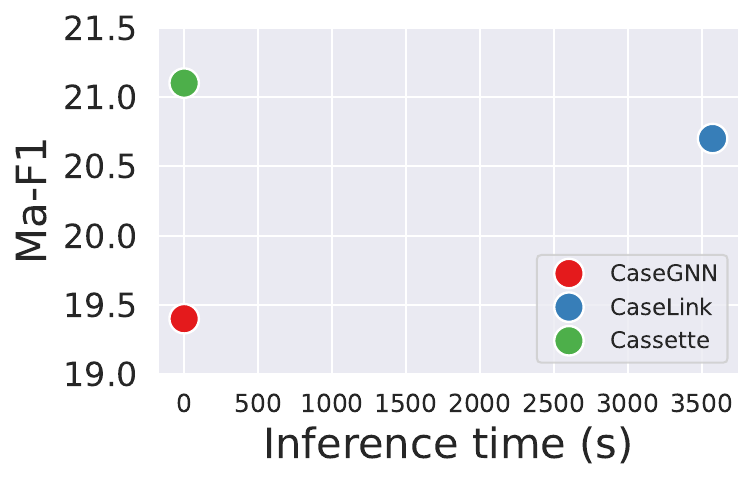}
    \label{fig:fig1_lecardv2}
    }
\caption{Retrieval performance versus inference efficiency of Cassette (\textcolor{ForestGreen}{green dots}), compared with two baselines: CaseGNN (\textcolor{red}{red dots}), an efficient but less effective two-tower method that servers as the initial feature, and CaseLink (\textcolor{NavyBlue}{blue dots}), a graph-based method that is more effective but computationally intensive. Results are shown on two datasets: (a) COLIEE2022 with 1,563 candidate cases (measured in milliseconds); and (b) LeCaRDv2 with 55,192 candidate cases (measured in seconds).}
\label{fig:motivation}
\end{figure}

Although graph-based LCR methods demonstrate strong retrieval accuracy, they face significant computational challenges in real-world applications. As illustrated in Figure~\ref{fig:motivation}, the heavy graph-based method CaseLink~\cite{caselink} (\textcolor{NavyBlue}{blue dots}) outperforms the traditional two-tower dense retrieval baseline, CaseGNN (\textcolor{red}{red dots}), in terms of accuracy. However, this performance gain comes at the cost of substantially reduced efficiency. \textbf{\textit{From the perspective of query-candidate case graph construction}}, establishing case-to-case relationships for incoming query cases is particularly time-consuming. In LCR task, query cases are unseen during model development, necessitating dynamic graph updates upon their arrival. This involves recalculating BM25-based edges and re-indexing~\cite{BM25}, resulting in \textbf{\textit{$O(n^2)$}} computational complexity due to the pairwise similarity computations ~\cite{caselink}. As shown in Figure~\ref{fig:motivation}, the graph-based model demonstrates an inference time of more than 500 milliseconds (ms) on the COLIEE2022\cite{COLIEE2022} dataset, which is approximately 50 times slower than the traditional retrieval baseline. On large scale datasets such as LeCaRDv2~\cite{lecardv2}, which includes 55,192 candidate cases, CaseLink exhibits extreme inference latency, which is approximately 340,000 times slower than the baseline method. This substantial delay renders graph-based approaches impractical for real-time retrieval scenarios, particularly when online processing is required for unseen queries in large legal databases.

\textbf{\textit{Additionally, from the perspective of case modelling with GNNs}}, the message passing operation in GNNs presents another major efficiency bottleneck due to the need for neighbourhood fetching and aggregation. In LCR scenario, retrieval is under an inductive setting during model testing, where candidate case pool is fixed while the query cases are unseen, similar to the legal practitioners searching for precedents for an unjudged case within a database of historically judged cases. For graph-based case-to-case retrieval methods, this inductive scenario necessitates dynamic reconstruction of the case graph and recalculation of case representations upon the arrival of each new query. Such requirements preclude offline pre-computation and leads to high online computational demands, highlighting the necessity for efficient LCR models that can maintain high retrieval accuracy while meeting the demands of real-world deployment.

\textbf{\textit{Moreover, considering the specific structural and semantic information in legal domain,}} directly distilling a MLP model in the context of LCR often results in suboptimal performance due to its limited capacity to capture and leverage rich legal information. In particular, the MLP encoder tends to disregard essential structural information, such as implicit case-to-case reference relationships, which is crucial for distinguishing legally relevant cases beyond surface-level textual similarity and for effective legal reasoning and retrieval. Cases with similar factual descriptions may lead to different judicial outcomes depending on factors such as applicable charges, precedent relationships, or shared legal issues. For example, cases involving similar facts but different charge types require retrieval systems to capture case–charge relationships rather than relying solely on surface-level textual similarity. Moreover, prior legal IR methods, such as SLR~\cite{SLR} and CFGL-LCR~\cite{cfgl}, have demonstrated that incorporating structural information within legal cases or modelling the structure relations among legal case graphs can significantly improve retrieval accuracy. Furthermore, naively applying distillation process focuses solely on transferring ranking signals from the teacher model to student model will result in failures of preserving the latent inter-case relationships encoded in the learned case representations. Therefore, an effective and ideal distillation strategy must consider both relevance ranking and structural alignment to ensure meaningful legal knowledge transfer.

In this paper, we propose a novel Cassette module by utilising distillation strategy involving ranking objective and eigen-matching objective to effectively transfer knowledge from a well-trained teacher retriever to a hybrid student dual encoder. First, a GNN-based teacher retriever is trained by leveraging the complete case-to-case relations including both query and candidate to capture case representative abilities. Subsequently, to increase the efficiency by avoiding the repetitive graph structure update, a GNN-based student candidate encoder is distilled to only calculate the case representations of candidate cases in an offline manner. To further relieve the computation overhead brought by GNN in online processing query cases, a MLP-based student query encoder is distilled to obtain effective query case representations. During distillation, a ranking objective and an eigen-matching objective are combined to distil case representation as well as the latent inter-case structural information from teacher retriever to hybrid student dual encoder. While existing approaches generally focus on model compression or the transfer of ranking supervision in generic retrieval settings, our method is tailored to legal case retrieval by distilling both semantic relevance and legally grounded structural relationships through a candidate legal graph. Unlike standard GNN-to-MLP distillation methods like Graph-MLP~\cite{graph-mlp} and GLNN~\cite{glnn}, Cassette does not aim to replicate the full graph reasoning process, but instead transfers structure-aware signals into a lightweight student suitable for large-scale deployment. Compared with conventional ranking distillation, our approach further incorporates legal-specific relational constraints, such as case–charge associations, making it more closely aligned with the unique requirements of the LCR task. This approach can also maintains the retrieval performance of the teacher model while significantly improving computational efficiency. 

In the proposed method Cassette, structurally guided distillation aligns with judicial reasoning principles because it transfers relational dependencies that reflect how courts analyse and compare cases. Judicial decision-making emphasizes consistency across similar charges, reliance on precedents, and adherence to structured legal constraints rather than isolated textual similarity. By distilling graph-based relational signals, the student model preserves proximity among cases sharing related charges or legal elements, promoting charge-level coherence in the representation space. At the same time, hierarchical relationships embedded in the legal graph enable the model to internalize precedent authority and analogical reasoning patterns. Unlike generic distillation methods that primarily transfer predictive distributions, our approach transfers legally structured inductive biases, ensuring that the learned representations are not only accurate but also aligned with core principles of judicial reasoning.

Experiments on three benchmark datasets, COLIEE2022~\cite{COLIEE2022}, COLIEE2023~\cite{COLIEE2023} and LeCaRDv2~\cite{lecardv2}, demonstrate the efficacy of the proposed distillation strategy, showcasing both its knowledge preservation capabilities and significant efficiency improvements. In addition to our previously proposed CaseLink~\cite{caselink}, the key contributions of this paper can be summarised as follows:
\begin{itemize} 
    \item A Cassette framework is proposed to distil efficient yet effective models for real-world LCR. Cassette includes a powerful but heavy GNN-based teacher retriever, and a lightweight but effective hybrid student dual encoder for query and candidate cases respectively. 
    \item A distillation strategy with a ranking objective and an eigen-matching objective are devised by transforming ranking results and graph structural information given by GNN-based teacher retriever into supervision signals for effective hybrid student dual encoder training.
    \item Extensive multi-lingual experiments on three benchmark datasets verify the outstanding performance and high efficiency of Cassette over the previous baselines.
\end{itemize}

\section{Related Work}
In this section, the recent progress in legal case retrieval, graph neural networks distillation and the ranking distillation will be briefly reviewed. 

\subsection{Legal Case Retrieval}
Legal case retrieval focuses on identifying relevant cases from a large database based on a query case, falling under the category of query-by-document tasks. There are recently two main branches of LCR models developed by the community: statistical models~\cite{TF-IDF,BM25,LMIR} and neural network models~\cite {Law2Vec,Lawformer,MTFT-BERT,MVCL,LEGAL-BERT,JOTR,DoSSIER,RPRS,NOWJ,UA@COLIEE2022,IOT-Match,Law-Match,LEDsummary,BM25injtct,LeiBi,LEVEN,JNLP@COLIEE2019,BERT-PLI,SAILER,promptcase,casegnn,caseencoder,gear,queryreformaulation,caselink,LawLLM,MileCut,uqlegalai,reakase,lexa}. Traditionally, the statistical models conduct the similarity search among the case databases with counting the term frequency with different strategies, including TF-IDF~\cite{TF-IDF}, BM25~\cite{BM25} and LMIR~\cite{LMIR}. For neural network models, a straightforward stream of methods is to make use of the language models such as BERT~\cite{BERT} to encode the case into high-dimensional vector with different strategies to deal with the extremely long cases against the fixed length input context of a language model, including truncating the first few words by SAILER~\cite{SAILER}, separating the case by paragraphs as in BERT-PLI~\cite{BERT-PLI}, by sentences as in IOT-Match~\cite{IOT-Match}, and by relational triplets as in CaseGNN~\cite{casegnn}. Recently, large language model-based methods such as LawLLM~\cite{LawLLM} and E5-Mistral~\cite{e5-mistral} are proposed to address different tasks in legal domain by unified fine-tuning strategy. Differently, the proposed Cassette can simultaneously achieve the efficiency of the two-tower retrieval and the efficacy of the relational modelling with graph structures.

Recently, given that GNNs are powerful tools for effective legal case representation learning, SOTA methods have been developed based on GNNs~\cite{LegalGNN,SLR,casegnn,caselink}. CaseGNN~\cite{casegnn} uses GNN to encode individual cases based on the relational triplet and retrieves relevant cases in the two-tower manner. SLR~\cite{SLR} and CFGL-LCR~\cite{cfgl} introduce GNNs into encoding the external knowledge graph to support LCR. Recently, CaseLink~\cite{caselink} develops a graph-based modelling of case-to-case relationship to extend the traditional two-tower manner with a more complex pairwise relational learning, which achieves the SOTA results. The proposed Cassette aims to preserve the outstanding performance of the complex method while increase the inference efficiency.

\subsection{Graph Neural Networks Distillation}
In the real-world application of graph neural networks~\cite{GCN,GAT,GraphSAGE,puma,trn-r1-zero,tntood}, the efficiency is always a critical aspect given that when the new data arrive, adding the new data node and edges into the existing graph and performing the message passing of GNN will be the main cost of the calculation~\cite{glnn,llp}.

In graph representation learning, knowledge distillation~\cite{kd} is effective to improve efficiency by training a small GNN from a large GNN, including TinyGNN~\cite{tinygnn}, LSP~\cite{lsp}, GFKD~\cite{gfkd}, KRD~\cite{krd}, CPF~\cite{cpf}, GraphAKD~\cite{graphakd}, FreeKD~\cite{freekd}, KDGA~\cite{kdga} and GKD~\cite{gkd}. Although these methods reduce the size of GNN models, the computational overhead related to the graph construction and the message passing operations are not improved.

More recently, GNN-to-MLP distillation has achieved a promising result in distilling the knowledge from GNN teacher model to MLP student model~\cite{glnn,llp,graph-mlp,nosmog,ff-g2m,adagmlp,vqgraph,tgs}. Most of the work focuses on distilling the structure-aware GNN models to a MLP model for node classification using general knowledge distillation objective, such as Graph-MLP~\cite{graph-mlp}, GLNN~\cite{glnn}, NOSMOG~\cite{nosmog}, VQGraph~\cite{vqgraph}, and TGS~\cite{tgs}. LLP~\cite{llp} applies similar distillation strategies to distil GNN to MLP for link prediction. The proposed Cassette roughly falls into this category yet it differs from the existing methods that Cassette primarily distils the ranking capability from the teacher GNN instead of the common classification ability.

\subsection{Ranking Distillation}
In information retrieval, a line of early research has explored the use of knowledge distillation to train lightweight retrievers from large, more powerful teacher models~\cite{rd,rd2,rankdistil,mo,rd-suite}. Notable examples include RD~\cite{rd} and RankDistil~\cite{rankdistil}, which adopt the general paradigm of knowledge distillation~\cite{kd} to transfer ranking knowledge from a high-capacity teacher to a smaller student model, typically within the same architectural family (e.g., two-tower models).

These approaches primarily aim to reduce inference latency while maintaining retrieval effectiveness by mimicking the teacher’s ranking behaviour. However, they generally assume that both teacher and student belong to the same class of models, thus limiting the flexibility and potential benefits of cross-architecture knowledge transfer.

In contrast, \textbf{Cassette} introduces a novel direction by performing \emph{cross-model} ranking distillation—transferring knowledge from a graph-based retriever (teacher) to a two-tower retriever (student). This paradigm shift enables the student to inherit the structural reasoning capabilities of the teacher while preserving the computational efficiency. To the best of our knowledge, Cassette is the first to explore this form of cross-model distillation in the legal case retrieval domain.

\section{Preliminary}
\subsection{Task Definition}
\label{sec:task}
In this work, we focus on a fixed candidate case pool setting, where the set of available cases remains constant throughout evaluation. During testing, only new queries are introduced, and the task is to retrieve the most relevant cases from this static pool. This setup reflects scenarios where the case repository is stable, and the system is evaluated on its ability to match unseen queries to existing cases. While our current experiments assume a fixed candidate case pool, the proposed framework is flexible and can be extended to handle incremental document insertion. In such a setting, new cases could be added to the pool over time, allowing the system to continuously update its retrieval capabilities. Exploring this dynamic scenario, however, lies beyond the scope of the current datasets and experiments, which are designed for fixed-pool evaluation.

With a set of $n$ candidate cases denoted as $\mathcal{D}=\{d_1,d_2,...,d_n\}$, and a set of $m$ query cases denoted as $\mathcal{Q}=\{q_1,q_2,...,q_m\}$, where each case $d$ could contain one or many legal charges $c\in\{c_1,c_2,...,c_o\}$ with $o$ charges native in the legal system, the task of legal case retrieval is to retrieve a set of relevant cases $\mathcal{D}^* = \{d^*_i| d^*_i \in \mathcal{D} \wedge \text{relevance}(d^*_i, q) \}$ from $\mathcal{D}$ with a given query case $q$. The function $\text{relevance}(d^*_i, q)$ represents that $d^*_i$ is a relevant case of query case $q$. Specifically, the relevant cases denote the precedents in legal domain, which are previous cases that was referred by another query case. In this paper, a case graph is denoted as $G = (\mathcal{V}, \mathcal{E})$, where $\mathcal{V}$ and $\mathcal{E}$ is the set of nodes and edges respectively in the case graph.

\subsection{Teacher Retriever Pretraining}
The teacher case retriever is obtained by pretraining a state-of-the-art graph-based CaseLink retriever~\cite{caselink}.

\paragraph{\textbf{Teacher Legal Graph Construction.}} The teacher legal graph (TLG) is constructed to include query cases, candidate cases and legal charges, with three intrinsic legal relations: case-to-case relations, case-to-charge relations and charge-to-charge relations. The overall adjacency matrix of TLG are set as undirected and unweighted, with the symmetric overall adjacency matrix, $\mathbf{A}_T \in \mathbb{R}^{(n+m+o)\times (n+m+o)}$ by combining the above edges:
\begin{equation}
 \mathbf{A}_T = \begin{bmatrix}
 \mathbf{A}_{T,d}  & \mathbf{A}_{T,b}^\intercal \\
 \mathbf{A}_{T,b} & \mathbf{A}_{T,c} 
 \end{bmatrix}.
\end{equation}
Case-Case edges in the TLG is denoted by $\mathbf{A}_{T,d} \in \mathbb{R}^{(n+m)\times (n+m)}$, where each value represents the BM25 similarity score between two cases in the training set. Since BM25 is not symmetric, simply choosing the larger one to represent the similarity between cases makes $\mathbf{A}_{T,d}$ symmetric. The Case-Charge edge exists when a charge name appears in a case text, which is denoted as $\mathbf{A}_{T,b}\in\mathbb{R}^{o\times (m+n)}$ and represented as binary entry. Charge-Charge edges are established by linking charge nodes with word embedding similarity and the adjacency matrix of Charge-Charge edges is denoted as $\mathbf{A}_{T,c}\in\mathbb{R}^{o\times o}$.

\paragraph{\textbf{Teacher Retriever Pretraining}}
\label{sec:teachergnnmodel}
With the constructed TLG, a GNN-based teacher CaseLink model parameterised by $\theta_T$ is developed to obtain case representations $\textbf{H}_T$ as:
\begin{equation}
\label{eq:gnnmodel}
    \textbf{H}_T=\text{GNN}_{\theta_T}(\textbf{X},\mathbf{A}_T),
\end{equation}
where $\mathbf{X}$ is the feature matrix consists of the node features (individual case or charge representations) $\mathbf{x}_q$,$\mathbf{x}_d,\mathbf{x}_c$ from language models. $\text{GNN}$ can be any graph neural networks, such as GCN~\cite{GCN}, GAT~\cite{GAT} or GraphSAGE~\cite{GraphSAGE}.

To train the teacher case retriever, a common practice in LCR task is to employ the contrastive learning with InfoNCE objective~\cite{infonce}:
\begin{equation}
\label{eq:infonce}
\ell_{\text{InfoNCE}}=-\text{log}\frac{e^{\text{Sim}(\mathbf{h}_{T,q},\mathbf{h}_{T,d^+})/\tau}}{e^{\text{Sim}(\mathbf{h}_{T,q},\mathbf{h}_{T,d^+})/\tau}+\sum^{p}_{i=1}e^{\text{Sim}(\mathbf{h}_{T,q},\mathbf{h}_{T,d^{-}_i})/\tau}},
\end{equation}
where Sim is a similarity measure function, one relevant (positive) cases $d^+$ and $p$ irrelevant (negative) cases $d^-$ are sampled for query case $q$, $\tau$ is the temperature parameter.

After pretraining, the teacher GNN will provide supervision signals to distil student encoders with frozen weights.

\section{Cassette Method}
With the powerful yet heavy teacher retriever, the aim of Cassette distillation framework is to obtain a model that can perform outstanding ranking with a much higher efficiency.

\begin{figure}[!t]
\centering
\includegraphics[width=1\linewidth]{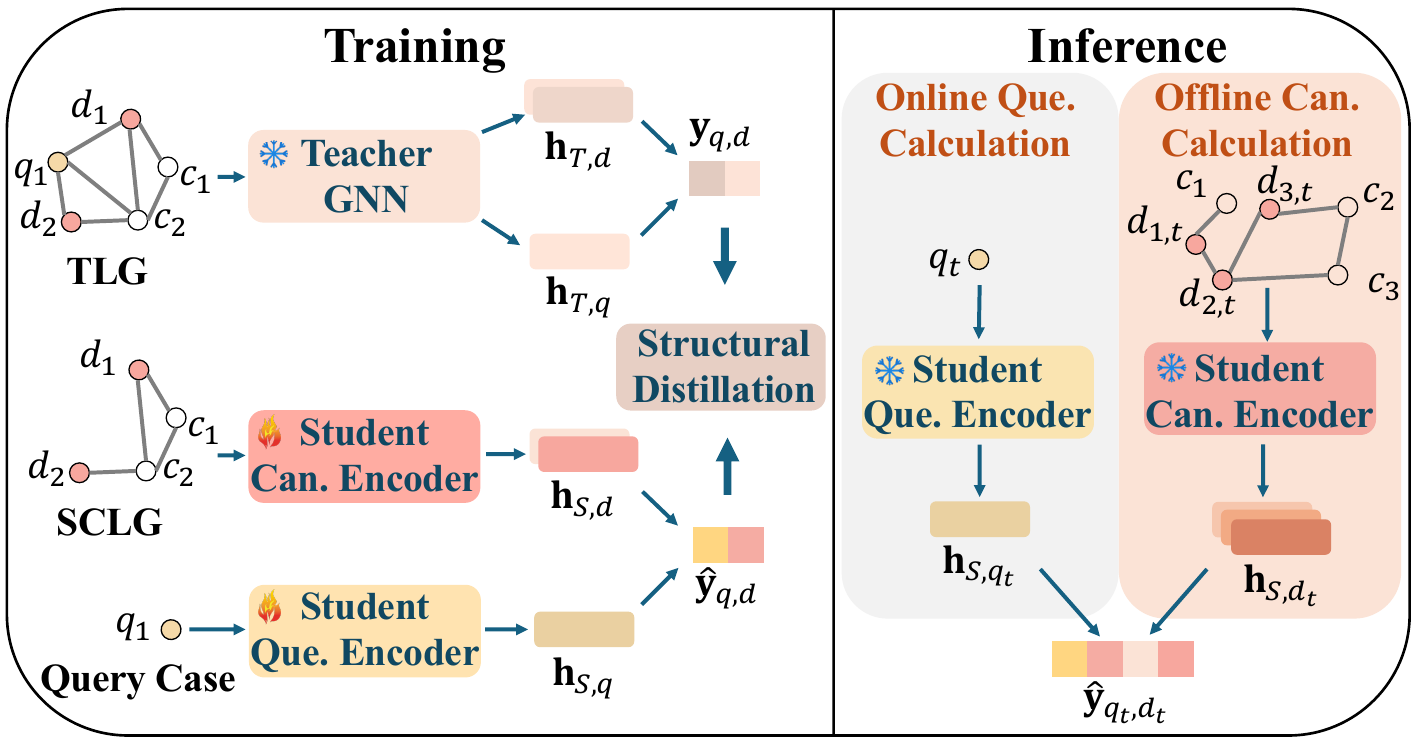}
\caption{Cassette consists of two phases: training and inference. On the left for training, a fixed, pretrained teacher GNN provides supervision signals to distil the knowledge into the student GNN for candidate (Can.) cases and the student MLP for query (Que.) cases. On the right for inference, the student GNN is used to obtain candidate case representations with offline calculation, while the student MLP processes incoming query cases online efficiently.}
\label{fig:Cassette}
\end{figure}

The design logic of Cassette is to avoid the dynamic update of the graph structure as in the GNN-based teacher retriever. When a new query case arrives to the LCR system, the Teacher Legal Graph will need to recalculate pairwise BM25 scores with new corpus, which further updates the graph structure $\mathbf{A}_T$. To prevent this dynamic and slow online calculation during model inference, a straightforward method would be simply distilling a two-tower model from teacher retriever. However, this strategy would lose the valuable structural information among cases. Therefore, \textbf{\textit{Cassette separates the modelling of candidate cases and query cases}} into dual student encoders by (1) constructing a static case graph for the candidate cases so that \textbf{\textit{the modelling of the candidates can be finished in offline}}; and (2) modelling the query cases with \textbf{\textit{a separate model for fast online calculation}}. An effective distillation strategy is needed to train the student encoders for legal case retrieval. The overall framework is illustrated in Figure~\ref{fig:Cassette}.

\subsection{Student Candidate Case Encoding}
Given that the pairwise case-case edges in the teacher model including both query and candidate cases, it is impractical to update the edges whenever a new query arrive. Generally, in legal case retrieval, query cases are unknown during the model development, while the candidate cases is stored in the database including all the precedent cases. Therefore, the candidate case representation can be designed to train in an effective graph-based encoder and pre-compute these representations offline.

\subsubsection{\textbf{Student Candidate Legal Graph}}
To avoid involving the unseen query case into the graph construction for efficient offline calculation, a student candidate legal graph (SCLG) is constructed only using candidate cases.

\textbf{Nodes}. SCLG includes candidate case nodes and charge nodes with input features encoded by language models, $\mathbf{x}_d$ and $\mathbf{x}_c \in \mathbb{R}^d$.

\textbf{Edges}. The edges in SCLG also includes three types: case-case edges, case-charge edges and charge-charge edges. 

\noindent$\bullet$ Case-Case Edge. In the candidate case graph, BM25 is utilised to calculate the similarity scores among candidate cases. The adjacency matrix of Case-Case edges, $\mathbf{A}_{S,d} \in \mathbb{R}^{n\times n}$ with $n$ candidate cases, in SCLG is calculated as:
    \begin{equation}
        \label{eq:d-edge_can}
        \mathbf{A}_{S,d_{ij}}=\left\{
        \begin{aligned}
        1 \quad & \text{for} & \text{TopK}(\text{BM25}(d_{i},d_{j}|d_{i},d_{j} \in \mathcal{D})). \\
        0 \quad & \text{for} & \text{Others}.
        \end{aligned}
        \right.
    \end{equation}
To ensure a symmetric $\mathbf{A}_{T,d}$ for stable calculation, if case $d_{i}$ appears in the top-k BM25 similarity of case $d_{j}$, yet $d_{j}$ does not appear in the top-k list of $d_{i}$, edges in both directions are still established. 

\noindent$\bullet$ Charge-Charge Edge. The adjacency matrix of charge-charge edges in student candidate case graph $\mathbf{A}_{S,c}\in\mathbb{R}^{o\times o}$ is denoted as:
\begin{equation}
        \label{eq:c-edge}
        \mathbf{A}_{S,c_{ij}}=\left\{
        \begin{aligned}
        1 \quad & \text{for} & \text{Sim}(\textbf{x}_{c_{i}},\textbf{x}_{c_{j}}|c_{i},c_{j} \in \mathcal{V})>\delta, \\
        0 \quad & \text{for} & \text{Others}, 
        \end{aligned}
        \right.
\end{equation}
where $c_{i}, c_{j}$ are two charge nodes in the set of node $\mathcal{V}$.

\noindent$\bullet$ Case-Charge Edge. The adjacency matrix of case-charge edges $\mathbf{A}_{S,b}\in\mathbb{R}^{o\times n}$ in SCLG is denoted as:
    \begin{equation}
        \label{eq:cd-edge_can}
        \mathbf{A}_{S,b_{ij}}=\left\{
        \begin{aligned}
        1 \quad & \text{for} & c_{i} \ \text{appears in} \ d_{j}|d_{j} \in \mathcal{D}, \\
        0 \quad & \text{for} & \text{Others},
        \end{aligned}
        \right.
    \end{equation}    

\noindent$\bullet$ The overall adjacency matrix of SCLG is symmetric, undirected and unweighted, $\mathbf{A}_S \in \mathbb{R}^{(n+o)\times (n+o)}$:
    \begin{equation}
     \mathbf{A}_S = \begin{bmatrix}
     \mathbf{A}_{S,d}  & \mathbf{A}_{S,b}^\intercal \\
     \mathbf{A}_{S,b} & \mathbf{A}_{S,c} 
     \end{bmatrix}.
    \end{equation}

\subsubsection{\textbf{Student Candidate Case Encoder}}
The student candidate encoder is developed similar to the teacher retriever in Eq.~\eqref{eq:gnnmodel}. Since the SCLG does not include query cases, the candidate cases in SCLG can be pre-computed with the student candidate encoder offline to improve the efficiency:
\begin{equation}
\label{eq:student_gnn}
    \textbf{H}_{S,d}=\text{GNN}_{\theta_S}(\textbf{X},\mathbf{A}_S),
\end{equation}
where $\mathbf{H}_{S,d}\in\mathbb{R}^{n\times k}$ are the representations of candidate cases omitting the charge representations, $\mathbf{h}_{S,d}$ is a candidate representation from $\mathbf{H}_S$, $\mathbf{X}$ includes candidate input features $\mathbf{x}_d$ and charge input features $\mathbf{x}_c$, and $\theta_S$ is the parameter of the student candidate encoder, which can be implemented as any graph neural networks, such as GCN~\cite{GCN}, GAT~\cite{GAT} or GraphSAGE~\cite{GraphSAGE}.

\subsection{Student Query Case Encoding}
Different from candidate cases, the newly incoming query case is encoded on the fly in Cassette with a multilayer perceptron (MLP) for the unseen query $q$ as:
    \begin{equation}
    \label{eq:student_mlp}
     \mathbf{h}_{S,q} = \text{MLP}(\mathbf{x}_{q}),
    \end{equation}
where $\mathbf{x}_{q}$ is the input representation of query $q \in \mathcal{D}$, $\mathbf{h}_{q}$ is the case representation of query $q$.

\subsection{Case to Case Structural Distillation}
To transfer the knowledge for legal case retrieval from the teacher retriever to the efficient student encoders, a structural distillation strategy is introduced to convert the ranking result of teacher model into supervision signals to train student models. Different from distillation for classification task where the objective can focus solely on distilling the logit from the classifier, the distillation for legal case retrieval task needs to transfer the query-candidate case ranking ability as well as the latent case-to-case relationship, which involves the structural relationship among legal cases. This fundamental difference necessitates the design of distinct distillation objectives. In the following, the teacher model $\text{GNN}_{\theta_T}$ is frozen during knowledge distillation process.          

\subsubsection{\textbf{Case Retrieval Ranking Matching}}
To retain the legal case retrieval ability from the teacher model, a ranking-based matching loss is devised. The key requirement for retrieval ability distillation is to correctly rank the candidate legal cases based on the query legal case. Given a query case $q$ and a candidate case $d$, the similarity score from the teacher legal case retriever and the score from the student encoders are denoted as $y_{q,d}$ and $\hat{y}_{q,d}$:
    \begin{equation}
    y_{q,d} = \text{Sim} (\mathbf{h}_{T,q}, \mathbf{h}_{T,d});\quad\hat{y}_{q,d} = \text{Sim} (\mathbf{h}_{S,q}, \mathbf{h}_{S,d}).
    \end{equation}
where $\mathbf{h}_{T,q}$ and $\mathbf{h}_{T,d}$ are the representations of query case $q$ and candidate case $d$ from the frozen teacher retriever in Eq.~\eqref{eq:gnnmodel}, $\mathbf{h}_{S,q}$ is the output representation of query $q$ from student query encoder in Eq.~\eqref{eq:student_mlp} and $\mathbf{h}_{S,d}$ denotes the output representation of candidate $d$ from the student candidate encoder in Eq.~\eqref{eq:student_gnn}. Sim refers to similarity functions that can measure the similarity between case representations, such as cosine similarity or dot product.

Given a set of candidate cases, the teacher similarity $y_{q,d}$ can be organised as a score list $\mathbf{y}_q$ for ranking, and similarly for the student similarity $\hat{y}_{q,d}$ to form a score list $\hat{\mathbf{y}}_q$. The ranking matching loss can be denoted as:
\begin{equation}
\label{eq:ranking_match}
    \ell_\text{ranking} = \frac{1}{m}\sum\limits_{q\in\mathcal{Q}}\text{Dist} (\mathbf{y}_q, \hat{\mathbf{y}}_q),
\end{equation}
where Dist is a distance function for measuring the distance between two ranking score list. The Dist function can be implemented with mean squared error (MSE) function, or Kullback-Leibler (KL) divergence with normalised ranking scores:
\begin{align}
    \ell_\text{ranking-MSE} &= \frac{1}{m}\sum\limits_{q\in\mathcal{Q}}||\mathbf{y}_q, \hat{\mathbf{y}}_q||^2_2,\\
    \ell_\text{ranking-KL} &= \frac{1}{m}\sum\limits_{q\in\mathcal{Q}}\text{KL} (\text{Softmax}(\mathbf{y}_q), \text{Softmax}(\hat{\mathbf{y}}_q)).
\end{align}

This distillation objective ensures the important query-candidate retrieval ability from the teacher model to provide supervision signals for the student model training to rank legal cases.

\subsubsection{\textbf{Case Eigen-Matching}}
In addition to focusing on query-candidate case ranking relationship in the ranking distillation objective, there are important latent structural relationships among legal cases missing by just distilling the query-candidate relationship. Therefore, a case eigen-matching objective is proposed to further distilling this structural information into the student model. For a graph, the eigenvalues and eigenvectors of the Laplacian matrix encode structural information of the graph. Smaller eigenvalues capture global structural patterns, such as connectivity or community structures, while larger eigenvalues capture local variations and fine-grained structural detail. Assume that a batch of legal cases, including $a$ query cases and $b$ candidate cases, are given to the teacher model. After case encoding, the case-to-case relationship can be obtained by measuring the pairwise similarity between case representations. Such latent pairwise relationship given by the teacher model can be denoted by a pseudo adjacency matrix:
    \begin{equation}
    \label{eq:teacher_adj}
    \mathbf{\bar A}_{T} = \begin{bmatrix}
     \text{Sim}(\mathbf{h}_{T,q_{1}},\mathbf{h}_{T,q_{1}}) &... & \text{Sim}(\mathbf{h}_{T,q_{1}},\mathbf{h}_{T,d_{b}}) \\
     \vdots & \ddots & \vdots \\    \text{Sim}(\mathbf{h}_{T,d_{b}},\mathbf{h}_{T,q_{1}}) &... & \text{Sim}(\mathbf{h}_{T,d_{b}},\mathbf{h}_{T,d_{b}}) \\
     \end{bmatrix},
    \end{equation}
where $\mathbf{\bar A}_{T} \in \mathbb{R}^{(a+b)\times(a+b)}$ and it is continuous. To mitigate the impact of negative similarity that are not meaningful in legal case relationships, and to ensure numerical stability in subsequent computations, all negative entries in $\mathbf{\bar A}_{T}$ are set to zero. This is equivalent to having no link between the nodes with negative similarity.

Similarly, with the same batch cases, the pseudo adjacency matrix $\mathbf{\bar A}_{S}\in\mathbb{R}^{(a+b)\times(a+b)}$ for student encoders, with negative entries set to zero, is denoted as:
    \begin{equation}
    \label{eq:student_adj}
    \mathbf{\bar A}_{S} = \begin{bmatrix}
     \text{Sim}(\mathbf{h}_{S,q_{1}},\mathbf{h}_{S,q_{1}}) &... & \text{Sim}(\mathbf{h}_{S,q_{1}},\mathbf{h}_{S,d_{b}}) \\
     \vdots & \ddots & \vdots \\    \text{Sim}(\mathbf{h}_{S,d_{b}},\mathbf{h}_{S,q_{1}}) &... & \text{Sim}(\mathbf{h}_{S,d_{b}},\mathbf{h}_{S,d_{b}}) \\
     \end{bmatrix}.
    \end{equation}
Specifically, Sim refers to similarity functions that can measure the similarity between case representations, such as cosine similarity or dot product. In Eq.~\eqref{eq:teacher_adj} and Eq.~\eqref{eq:student_adj}, cosine similarity is chosen as the similarity function.

To utilise the eigen-characteristics to distil the learned knowledge from the teacher model, the pseudo adjacency matrices $\mathbf{\bar A}_{T}$ and $\mathbf{\bar A}_{S}$ are converted into their corresponding normalized Laplacian matrices as: (1) $\mathbf{L}_{T} = \mathbf{I}-\mathbf{D}_{T}^{-1/2}\mathbf{\bar A}_{T}\mathbf{D}_{T}^{-1/2}$, where $\mathbf{I}$ is an identity matrix and $\mathbf{D}_{T}$ is the degree matrix. The resulting normalizes Laplacian matrix is then decomposed into $\mathbf{L}_{T} = \mathbf{U} \mathbf{\Lambda}_{T} \mathbf{U}^{T}$, where $\mathbf{\Lambda}_{T}$ is the diagonal eigenvalue matrix for representations of the teacher model. (2) Similarly, $\mathbf{L}_{S} = \mathbf{I}-\mathbf{D}_{S}^{-1/2}\mathbf{\bar A}_{S}\mathbf{D}_{S}^{-1/2}$. And the decomposition is $\mathbf{L}_{S} = \mathbf{U} \mathbf{\Lambda}_{S} \mathbf{U}^{T}$, where $\mathbf{\Lambda}_{S}$ is the eigenvalue matrix.

With the eigenvalues of the pseudo adjacency matrices from the teacher retriever and student encoders, the eigen-matching loss is given by L2 distance with $\mathcal{C}$ batches of legal cases ($|\mathcal{C}|=\lceil m/a\rceil$):
\begin{equation}
\label{eq:eig_match}
    \ell_\text{eigen} =\frac{1}{|\mathcal{C}|}\sum\limits_{\mathcal{C}}\frac{1}{a+b}||\mathbf{\Lambda}_{T}-\mathbf{\Lambda}_{S}||^2_{2}.
\end{equation}
The reason why eigenvalue matrices can be directly deducted is based on the fact that the original matrices are aligned in terms of nodes (legal cases). The distillation objective is designed to preserve the structural information encoded in the teacher model's latent case representations, enhancing the student's ability to capture and utilise structural patterns of legal cases.

\subsubsection{\textbf{Supervised Loss}}
To facilitate case retrieval, the commonly used contrastive learning objective $\ell_{\text{InfoNCE}}$ (Eq.~\eqref{eq:infonce}) is used as the supervised loss $\ell_{\text{sup}}$ to further enhance the retrieval capability of the hybrid student dual encoder. During the ranking distillation process, positive samples are derived from the ground truth labels, while easy negative samples are randomly selected from the case pool $\mathcal{D}$. 
To guide training with harder samples, hard negatives are selected based on the BM25~\cite{BM25} relevance score. Cases with high BM25 scores but not a label case are chosen as hard negatives.

\subsubsection{Overall Student Distillation Objective}
The overall objective for student GNN and student MLP ranking distillation is:
\begin{equation}
\label{eq:loss-overall}
    \ell_\text{Student} = \alpha\cdot\ell_\text{ranking}+\beta\cdot\ell_\text{eigen}+\mu\cdot\ell_{\text{sup}},
\end{equation}
where $\alpha$, $\beta$ and $\mu$ are the coefficients that regulate the relative weight of three loss functions.

\subsection{Inference}
During student dual encoder inference phase, given a query in the test query set $q_{\text{t}}\in \mathcal{Q}_{\text{t}}$ and a candidate in test candidate set $d_{\text{t}} \in \mathcal{D}_{\text{t}}$, the similarity score $\hat{y}_{q_{\text{t}},d_{\text{t}}}$ is computed as:
\begin{equation} 
\label{eq:inference}
    \hat{y}_{q_{\text{t}},d_{\text{t}}} = \text{Sim} (\mathbf{h}_{S,q_{\text{t}}}, \mathbf{h}_{S,d_{\text{t}}}),
\end{equation}
where $\mathbf{h}_{S,q_{\text{t}}}$ is the representation of query $q$ from student query encoder and $\mathbf{h}_{S,d_{\text{t}}}$ is the representations of candidate $d_{\text{t}}$ from student candidate encoder. Candidates with top ranking scores are retrieved. The overall framework is illustrated in Figure~\ref{fig:Cassette}.

The reason of using student candidate case encoder instead of teacher retriever for candidates encoding is because the teacher retriever is trained to operate on a graph with both candidate cases and query cases, but in the testing graph, only candidate cases are available. If the teacher retriever is applied during testing, the performance will drop significantly. The comparison between teacher retriever and student candidate encoder for testing is detailed in Section~\ref{sec:comparisonteachernstudentgnn}.

It is worth noting that, although the LCR task is inductive — there is no overlap between the training and test sets, the proposed Cassette model \textbf{\textit{does not need retraining during testing}} when the database is expanding. This is achieved by simply connecting test candidates into the offline Student Candidate Legal Graph and directly performing inference using the trained student candidate encoder. Moreover, the scenario in which new query case becomes a candidate in the candidate pool for next query is not considered in LCR task setting in this paper. Such a retraining-free approach offers an efficient and effective solution.

\section{Experiment}
In this section, the experiment settings and results are described to answer the following research questions (RQs):
\begin{itemize}
    \item RQ1: How does Cassette perform compared with baselines?
    \item RQ2: How efficient is Cassette compared with baselines?
    \item RQ3: How effective is each module in Cassette?
    \item RQ4: How does ranking and eigen-matching distillation help with Cassette?
    \item RQ5: How effective is the student candidate case encoder?
    \item RQ6: How effectiveness of incremental document insertion without model retraining? 
    \item RQ7: How robust is the model with respect to variations in charge categories?
    \item RQ8: How do hyper-parameter settings affect Cassette?
\end{itemize}

\subsection{Setup}
\label{sec:setup}
\begin{table}[!t]\centering
\caption{Statistics of datasets.}\label{tab:dataset}
\resizebox{!}{!}{
    \begin{tabular}{c|cc|cc|cc}
    \toprule
    \multirow{2}{*}{Datasets} &\multicolumn{2}{c|}{COLIEE2022} &\multicolumn{2}{c|}{COLIEE2023} &\multicolumn{2}{c}{LeCaRDv2} \\
    \cmidrule{2-7}
    &train &test &train &test &train &test \\\midrule
    Language &\multicolumn{2}{c|}{English} &\multicolumn{2}{c|}{English} &\multicolumn{2}{c}{Chinese} \\
    \# Query &898 &300 &959 &319 &640 &160\\
    \# Candidates &4415 &1563 &4400 &1335 &\multicolumn{2}{c}{55192}\\
    \# Avg. relevant cases &4.68 &4.21 &4.68 &2.69 &23.79 &24.35\\
    Avg. length (\# token) &6724 &6785 &6532 &5566 &\multicolumn{2}{c}{4766}\\
    Largest length (\# token) &127934 &85136 &127934 &61965 &\multicolumn{2}{c}{222069} \\
    \bottomrule
    \end{tabular}
    }
\end{table}

\subsubsection{Datasets.}
The experiments are conducted on three benchmark datasets with statistics in Table~\ref{tab:dataset}.
\begin{itemize}
    \item \textbf{COLIEE2022}~\cite{COLIEE2022} and \textbf{COLIEE2023}~\cite{COLIEE2023} are two English datasets from the Competition on Legal Information Extraction/Entailment (COLIEE) and consist of cases from the Federal Court of Canada. There are two key differences between them. First, the cases in the test sets and the majority of the training sets are different. Second, as shown in Table~\ref{tab:dataset}, the average number of relevant cases per query is different, which leads to different levels of difficulty. 
    \item \textbf{LeCaRDv2} is a Chinese dataset collected from the China Judgment Online, which consists of 800 queries and 55,192 candidates of criminal cases~\cite{lecardv2}. Different from the totally separate training set and test set in COLIEE, both the training candidates and test candidates come from the same candidate case pool in LeCaRDv2. Yet the training and test query cases are different, and test queries are unseen during training, which is still an inductive retrieval setting.
\end{itemize}

\subsubsection{Metrics.}
\label{metrics}
To evaluate the performance, we used the following metrics: precision (P), recall (R), Micro F1 (Mi-F1), Macro F1 (Ma-F1), Mean Reciprocal Rank (MRR), Mean Average Precision (MAP), and normalized discounted cumulative gain (NDCG), as these are widely employed in information retrieval tasks. If not explicitly specified, Top-5 results were reported for COLIEE and Top-30 for LeCaRDv2 following previous LCR methods~\cite{promptcase,LeCaRD,SAILER,casegnn,caselink,lecardv2}. For all metrics, higher values indicate better performance.

\subsubsection{Baselines.}
\label{baselines}
To assess the performance of Cassette, we conducted experiments comparisons with the following baselines:
\begin{itemize}
    \item \textbf{BM25}~\cite{BM25}: a traditional yet robust retrieval benchmark that measures text similarity using term frequency.
    \item \textbf{\textcolor{purple}{Language model (LM)-based methods}}:
    \item \textbf{LEGAL-BERT (2020)}~\cite{LEGAL-BERT}: a LM pretrained on a large English legal corpus on a traditional BERT~\cite{BERT} model.
    \item \textbf{Lawformer (2021)}~\cite{Lawformer}: a longformer-based pre-trained language model for Chinese legal long documents understanding and encoding.
    \item \textbf{MonoT5 (2019)}~\cite{monot5}: a sequence-to-sequence document ranking model based on the T5 architecture.
    \item \textbf{SAILER (2023)}~\cite{SAILER}: a legal structure-aware model that achieves competitive results on the same datasets.
    \item \textbf{PromptCase (2023)}~\cite{promptcase}: a prompt-based input reformulation method for legal case retrieval (LCR).
    \item \textbf{\textcolor{purple}{Large language model (LLM)-based methods}}:
    \item \textbf{E5-Mistral-7b-Instruct (2024)}~\cite{e5-mistral}: a large language model (LLM) that achieved top-ranking performance in legal retrieval tasks on the MTEB~\cite{mteb} benchmark.~\cite{mteb}\footnote{The evaluation of MTEB is different from the official dataset evaluation that MTEB's evaluation is based on sampling ranking while the official dataset evaluation is based on full candidate set ranking. This paper follows the official dataset evaluation to keep consistency with other LCR baselines.}.
    \item \textbf{Qwen3-Embedding-8B (2025)}~\cite{qwen3embedding}: a large embedding model from the Qwen series that produces robust, versatile vector representations optimized for retrieval and semantic similarity, with strong performance across diverse domains.
    \item \textbf{Inf-Retriever-V1 (2025)}~\cite{inf-retriever-v1}: an LLM-based dense retrieval model, achieving state-of-the-art performance on legal retrieval tasks in MTEB by generating highly discriminative embeddings that enhance both accuracy and scalability in downstream applications.
    \item \textbf{\textcolor{purple}{Graph-based methods}}:
    \item \textbf{CaseGNN (2024)}~\cite{casegnn}: a two-tower method by encoding the entities of individual cases with GNN. Yet there is no Chinese-based variant available for LeCaRDv2 dataset.
    \item \textbf{CaseLink (2024)}~\cite{caselink}: a SOTA method based on GNN with case-to-case relationship, which is the teacher GNN.
\end{itemize}

\subsubsection{Implementation.}
\label{implementation}
The default ranking distillation objective is set to MSE. The batch size of training teacher retriever is chosen from \{32, 64, 128\} while the batch size of training dual student encoders are chosen from \{32, 64, 128, 256, 512, \#Query\}. GAT~\cite{GAT} is the default GNN for teacher and student models with number of layers chosen from \{1,2,3\}. The number of layers of the student query encoder is chosen from \{2, 3, 4\}. The hidden dimension is set to 1536 as the size of input features. The dropout~\cite{dropout} rate for teacher and student models are chosen from \{0, 0.1, 0.2, 0.3, 0.4, 0.5\}. Adam~\cite{Adam} is the default optimiser with the learning rate chosen from \{1e-3, 5e-4, 1e-4, 5e-5, 1e-5\} and the weight decay \{1e-4, 1e-5, 1e-6\}. For the supervised contrastive learning, the number of positive is $1$ while the number of easy negative and hard negative are chosen from \{5, 10\}. CaseGNN~\cite{casegnn} serves as input feature encoder and SAILER~\cite{SAILER} is utilised as the charge encoder for COLIEE datasets. E5-Mistral-7b-Instruct~\cite{e5-mistral} serves as input feature and charge encoder for LeCaRDv2. The two-stage experiment is conducted using top 10 and top 50 BM25 ranking results for COLIEE and LeCaRDv2, separately. The results of the two-stage experiment are reported solely for overall comparison to verify LCR performance and all other experiments report only one-stage results. The number of top K case neighbour nodes in Eq.~\eqref{eq:d-edge_can} is selected from {3, 5, 10, 30, 40}, and the threshold $\delta$ in Eq.~\eqref{eq:c-edge} is chosen from {0.85, 0.9, 0.95}. in Eq.~\eqref{eq:loss-overall}, the coefficients $\alpha$ $\mu$ and $\beta$ are chosen from \{0, 0.2, 0.5, 0.8, 1\}. The French text in COLIEE datasets, which only appears in a small portion of cases from Quebec, Canada, was removed to ensure consistent text encoding. The evaluation follows an inductive setting, where all queries in the test set are unseen during training.

\begin{table*}[!t]\centering
\caption{Overall performance on COLIEE2022 (\%). Underlined numbers indicate the best baselines. Bold numbers indicate the best performance of all methods. Both one-stage and two-stage Top-5 results are reported. $\Delta$ Init represents the performance gap between the feature initialization model, CaseGNN, and Cassette. $\Delta$ GNN denotes the performance gap between SOTA GNN-based model, CaseLink, and Cassette.}\label{tab:cassette_overall_2022}
\resizebox{\linewidth}{!}{
\begin{tabular}{c|l|ccccccc}
\toprule
\midrule
&\multirow{2}{*}{Methods} &\multicolumn{7}{c}{COLIEE2022} \\
\cmidrule{3-9}
&&P@5 &R@5 &Mi-F1 &Ma-F1 &MRR@5 &MAP &NDCG@5 \\\midrule
\midrule
&\textbf{One-stage}\\
&BM25 &17.9 &21.2 &19.4 &21.4 &23.6 &25.4 &33.6\\
\midrule
\multirow{4}{*}{\rotatebox{90}{LM}}&LEGAL-BERT &4.47 &5.30 &4.85 &5.38 &7.42 &7.47 &10.9\\%\midrule
&MonoT5 &0.71 &0.65 &0.60 &0.79 &1.39 &1.41 &1.73\\%\midrule
&SAILER &16.6 &15.2 &14.0 &16.8 &17.2 &18.5 &25.1\\%\midrule
&PromptCase &17.1 &20.3 &18.5 &20.5 &35.1 &33.9 &38.7\\
\midrule
\multirow{3}{*}{\rotatebox{90}{LLM}}&E5-Mistral-7b-Instruct &21.4 &25.4 &23.2 &25.7 &26.8 &28.5 &38.0\\%\midrule
&Qwen3-Embedding-8B &21.6 &25.7 &23.5 &26.0 &26.7 &29.0 &37.8\\%\midrule
&Inf-Retriever-V1 &21.1 &25.1 &22.9 &25.5 &26.6 &28.7 &37.9 \\
\midrule
\multirow{3}{*}{\rotatebox{90}{Graph}}&CaseGNN&35.5 $\pm$0.2 &42.1 $\pm$0.2 &38.4 $\pm$0.3 &42.4 $\pm$0.1 &66.8 $\pm$0.8 &64.4 $\pm$0.9 &69.3 $\pm$0.8\\
&CaseLink &\underline{37.0} $\pm$0.1&\underline{43.9} $\pm$0.1&\underline{40.1} $\pm$0.1&\underline{44.2} $\pm$0.1&\underline{\textbf{67.3}} $\pm$0.5&\underline{\textbf{65.0}} $\pm$0.2&\underline{\textbf{70.3}} $\pm$0.1\\
&\cellcolor{lightgray}Cassette (Ours) &\cellcolor{lightgray}\textbf{37.2} $\pm$0.1 &\cellcolor{lightgray}\textbf{44.2} $\pm$0.2 &\cellcolor{lightgray}\textbf{40.4} $\pm$0.1 &\cellcolor{lightgray}\textbf{44.6} $\pm$0.1 &\cellcolor{lightgray}66.9 $\pm$0.1 &\cellcolor{lightgray}64.4 $\pm$0.1 &\cellcolor{lightgray}69.7 $\pm$0.1\\
\midrule
&$\Delta$ Init &+1.7 &+2.1 &+2.0 &+2.2 &+0.1 &+0.0 &+0.5\\
&$\Delta$ GNN &+0.2 &+0.3 &+0.3 &+0.4 &-0.4 &-0.6 &-0.6\\
\midrule
\midrule
&\textbf{Two-stage} \\
\multirow{2}{*}{\rotatebox{90}{LM}}&SAILER &23.8 &25.7 &24.7 &25.2 &43.9 &42.7 &48.4 \\%\midrule
&PromptCase &23.5 &25.3 &24.4 &\underline{\textbf{30.3}} &41.2 &39.6 &45.1\\
\midrule
\multirow{3}{*}{\rotatebox{90}{LLM}}&E5-Mistral-7b-Instruct &22.0 &26.2 &24.0 &25.8 &48.4 &46.8 &51.5 \\
&Qwen3-Embedding-8B &21.9 &26.0 &23.8 &25.8 &48.8 &48.0 &52.1\\
&Inf-Retriever-V1 &22.2 &26.4 &24.1 &26.1 &45.5 &44.5 &49.6\\
\midrule
\multirow{3}{*}{\rotatebox{90}{Graph}}&CaseGNN &22.9 $\pm$0.1 &27.2 $\pm$0.1 &24.9 $\pm$0.1 &27.0 $\pm$0.1 &54.9 $\pm$0.4 &54.0 $\pm$0.5 &57.3 $\pm$0.6\\
&CaseLink &\underline{24.7} $\pm$0.1 &\underline{29.1} $\pm$0.1 &\underline{26.8} $\pm$0.1 &29.2 $\pm$0.1 &\underline{56.0} $\pm$0.2 &\underline{55.0} $\pm$0.2 &\underline{58.6} $\pm$0.1\\
&\cellcolor{lightgray}Cassette (Ours) &\cellcolor{lightgray}\textbf{24.9} $\pm$0.1 &\cellcolor{lightgray}\textbf{29.6} $\pm$0.1 &\cellcolor{lightgray}\textbf{27.1} $\pm$0.1 &\cellcolor{lightgray}29.4 $\pm$0.1 &\cellcolor{lightgray}\textbf{56.7} $\pm$0.1 &\cellcolor{lightgray}\textbf{55.7} $\pm$0.1 &\cellcolor{lightgray}\textbf{59.1} $\pm$0.1\\
\midrule
\multirow{2}{*}{\rotatebox{90}{}}&$\Delta$ Init &+2.0&+2.4&+2.2&+2.4&+1.8&+1.7&+1.8\\
&$\Delta$ GNN &+0.2&+0.5&+0.3&+0.2&+0.7&+0.7&+0.5\\
\bottomrule
\end{tabular}}
\end{table*}

\begin{table*}[!t]\centering
\caption{Overall performance on COLIEE2023 (\%). Underlined numbers indicate the best baselines. Bold numbers indicate the best performance of all methods. Both one-stage and two-stage Top-5 results are reported. $\Delta$ Init represents the performance gap between the feature initialization model, CaseGNN, and Cassette. $\Delta$ GNN denotes the performance gap between SOTA GNN-based model, CaseLink, and Cassette.}\label{tab:cassette_overall_2023}
\resizebox{\linewidth}{!}{
\begin{tabular}{c|l|cccccccc}
\toprule
\midrule
&\multirow{2}{*}{Methods} &\multicolumn{7}{c}{COLIEE2023} \\
\cmidrule{3-9}
&&P@5 &R@5 &Mi-F1 &Ma-F1 &MRR@5 &MAP &NDCG@5 \\\midrule
\midrule
\multirow{2}{*}{\rotatebox{90}{ }}&\textbf{One-stage}\\
&BM25 &16.5 &30.6 &21.4 &22.2 &23.1 &20.4 &23.7\\
\midrule
\multirow{4}{*}{\rotatebox{90}{LM}} &LEGAL-BERT &4.64 &8.61 &6.03 &6.03 &11.4 &11.3 &13.6\\%\midrule
&MonoT5 &0.38 &0.70 &0.49 &0.47 &1.17 &1.33 &0.61 \\%\midrule
&SAILER &12.8 &23.7 &16.6 &17.0 &25.9 &25.3 &29.3\\%\midrule
&PromptCase &16.0 &29.7 &20.8 &21.5 &32.7 &32.0 &36.2 \\%\midrule
\midrule
\multirow{3}{*}{\rotatebox{90}{LLM}}&E5-Mistral-7b-Instruct &16.0 &29.7 &20.8 &21.5 &21.9 &22.9 &30.8\\
&Qwen3-Embedding-8B &18.4 &34.1 &23.9 &25.2 &25.8 &27.5 &36.4\\
&Inf-Retriever-V1 &19.0 &35.3 &24.7 &26.0 &26.5 &28.0 &37.6 \\
\midrule
\multirow{3}{*}{\rotatebox{90}{Graph}}&CaseGNN &17.7 $\pm$0.7 &32.8 $\pm$0.7 &23.0 $\pm$0.5 &23.6 $\pm$0.5 &38.9 $\pm$1.1 &37.7 $\pm$0.8 &42.8 $\pm$0.7\\
&CaseLink &\underline{\textbf{20.9}} $\pm$0.3&\underline{\textbf{38.4}} $\pm$0.6&\underline{\textbf{27.1}} $\pm$0.3&\underline{\textbf{28.2}} $\pm$0.3&\underline{\textbf{45.8}} $\pm$0.5&\underline{\textbf{44.3}} $\pm$0.7&\underline{\textbf{49.8}} $\pm$0.4\\
&\cellcolor{lightgray}Cassette (Ours) &\cellcolor{lightgray}19.8 $\pm$0.1 &\cellcolor{lightgray}36.7 $\pm$0.2 &\cellcolor{lightgray}25.7 $\pm$0.1 &\cellcolor{lightgray}26.6 $\pm$0.2 &\cellcolor{lightgray}42.5 $\pm$0.2 &\cellcolor{lightgray}41.1 $\pm$0.1 &\cellcolor{lightgray}46.8 $\pm$0.2\\
\midrule
\multirow{2}{*}{\rotatebox{90}{}}&$\Delta$ Init &+2.1&+3.9&+2.7&+3.0&+3.6&+3.4&+4.0\\
&$\Delta$ GNN &-1.1&-1.7&-1.4&-1.6&-3.3&-3.2&-3.0\\
\midrule
\midrule
&\textbf{Two-stage} \\
\multirow{2}{*}{\rotatebox{90}{LM}} &SAILER &19.6 &32.6 &24.5 &23.5 &37.3 &36.1 &40.8\\
&PromptCase &\underline{21.8} &36.3 &\underline{27.2} &26.5 &39.9 &38.7 &44.0\\
\midrule
\multirow{3}{*}{\rotatebox{90}{LLM}} &E5-Mistral-7b-Instruct &19.7 &36.6 &25.6 &26.5 &46.9 &44.4 &49.9\\
&Qwen3-Embedding-8B &20.1 &37.4 &26.2 &27.0 &47.7 &46.3 &51.1\\
&Inf-Retriever-V1 &19.7 &36.6 &25.6 &26.6 &48.2 &46.9 &51.6\\
\midrule
\multirow{3}{*}{\rotatebox{90}{Graph}} &CaseGNN &20.2 $\pm$0.2 &37.6 $\pm$0.5 &26.3 $\pm$0.3 &27.3 $\pm$0.2 &45.8 $\pm$0.9 &44.4 $\pm$0.8 &49.6 $\pm$0.8\\
&CaseLink &21.0 $\pm$0.3 &\underline{38.9} $\pm$0.5 &27.1 $\pm$0.3 &\underline{28.2} $\pm$0.3 &\underline{48.8} $\pm$0.2 &\underline{47.2} $\pm$0.1 &\underline{52.6} $\pm$0.1\\
&\cellcolor{lightgray}Cassette (Ours) &\cellcolor{lightgray}\textbf{22.1} $\pm$0.1 &\cellcolor{lightgray}\textbf{41.0} $\pm$0.1 &\cellcolor{lightgray}\textbf{28.7} $\pm$0.1 &\cellcolor{lightgray}\textbf{29.6} $\pm$0.1 &\cellcolor{lightgray}\textbf{49.7} $\pm$0.1 &\cellcolor{lightgray}\textbf{48.2} $\pm$0.2 &\cellcolor{lightgray}\textbf{53.7} $\pm$0.1\\
\midrule
\multirow{2}{*}{\rotatebox{90}{}}&$\Delta$ Init &+1.9&+3.4&+2.4&+2.3&+3.9&+3.8&+4.1\\
&$\Delta$ GNN &+0.9&+1.1&+1.6&+1.4&+0.9&+1.0&+1.1\\
\bottomrule
\end{tabular}}
\end{table*}

\begin{table}[!t]\centering
\caption{Overall performance on LeCaRDv2 (\%). Underlined numbers indicate the best baselines. Bold numbers indicate the best performance of all methods. Both one-stage and two-stage Top-30 results are reported. $\Delta$ Init represents the performance gap between the feature initialization model, E5-Mistral-7b-Instruct, and Cassette. $\Delta$ GNN denotes the performance gap between SOTA GNN-based model, CaseLink, and Cassette.}\label{tab:overall_lecardv2}
\resizebox{!}{!}
{
\begin{tabular}{c|l|cccccccc}
\toprule \midrule
&\multirow{2}{*}{Methods} &\multicolumn{7}{c}{LeCaRDv2} \\
\cmidrule{3-9}
&&P@30 &R@30 &Mi-F1 &Ma-F1 &MRR@30 &MAP &NDCG@30 \\\midrule
\midrule
\multirow{2}{*}{\rotatebox{90}{ }}&\textbf{One-stage}\\
&BM25 &19.3 &23.6 &21.2 &20.7 &31.2 &33.7 &50.8  \\\midrule
\multirow{3}{*}{\rotatebox{90}{LM}} &SAILER &11.3 &13.9 &12.5 &12.6 &43.5 &31.0 &48.8 \\%\midrule
&PromptCase &13.9 &17.0 &15.3 &15.3 &51.3 &35.9 &54.6 \\%\midrule
&Lawformer &9.45 &11.6 &10.4 &9.86 &38.6 &32.9 &42.2\\\midrule
\multirow{3}{*}{\rotatebox{90}{LLM}}&E5-Mistral-7b-Instruct &18.0 &22.0 &19.8 &19.4 &57.8 &40.5 &59.1\\%\midrule
&Qwen3-Embedding-8B &16.7 &20.5 &18.4 &17.9 &59.0 &43.3 &59.6 \\
&Inf-Retriever-V1 &15.4 &18.8 &16.9 &16.3 &53.4 &41.9 &56.7\\\midrule
\multirow{2}{*}{\rotatebox{90}{Graph}}&CaseLink &19.1$\pm$0.1 &23.3$\pm$0.1 &21.0$\pm$0.3 &20.7$\pm$0.1 &63.3$\pm$0.3 &44.0$\pm$0.2 &62.3$\pm$0.4 \\
&\cellcolor{lightgray}Cassette (Ours) &\cellcolor{lightgray}\textbf{19.5}$\pm$0.1 &\cellcolor{lightgray}\textbf{23.9}$\pm$0.1 &\cellcolor{lightgray}\textbf{21.4}$\pm$0.1 &\cellcolor{lightgray}\textbf{21.1}$\pm$0.1 &\cellcolor{lightgray}\textbf{63.7}$\pm$0.9 &\cellcolor{lightgray}\textbf{45.0}$\pm$0.4 &\cellcolor{lightgray}\textbf{63.4}$\pm$0.3\\
\midrule
\multirow{2}{*}{\rotatebox{90}{ }}&$\Delta$ Init &+1.5 &+1.9 &+1.6 &+1.7 &+5.9 &+4.5 &+4.3\\
&$\Delta$ GNN &+0.4 &+0.6 &+0.4 &+0.4 &+0.4 &+1.0 &+1.1 \\
\midrule
\midrule
&\textbf{Two-stage} \\
\multirow{3}{*}{\rotatebox{90}{LM}} &SAILER &19.7 &24.2 &21.8 &21.3 &57.5 &42.1 &60.2 \\
&PromptCase &20.3 &24.8 &22.3 &21.7 &58.9 &44.5 &61.9 \\
&Lawformer &19.2 &23.5 &21.1 &20.4 &55.8 &41.6 &57.4 \\\midrule
\multirow{3}{*}{\rotatebox{90}{LLM}}&E5-Mistral-7b-Instruct &21.3 &26.1 &23.5 &22.9 &64.7 &49.9 &65.6 \\
&Qwen3-Embedding-8B &20.7 &25.4 &22.8 &22.3 &64.1 &48.3 &64.5 \\
&Inf-Retriever-V1 &20.0 &24.4 &22.0 &21.3 &61.2 &46.4 &62.6\\\midrule
\multirow{2}{*}{\rotatebox{90}{Graph}}
&CaseLink &21.4$\pm$0.1 &26.2$\pm$0.2 &23.6$\pm$0.1 &23.0$\pm$0.1 &66.0$\pm$0.6 &50.2$\pm$0.1 &66.0$\pm$0.2\\
&\cellcolor{lightgray}Cassette (Ours) &\cellcolor{lightgray}\textbf{21.6}$\pm$0.1 &\cellcolor{lightgray}\textbf{26.5}$\pm$0.1 &\cellcolor{lightgray}\textbf{23.8}$\pm$0.1 &\cellcolor{lightgray}\textbf{23.2}$\pm$0.1 &\cellcolor{lightgray}\textbf{67.2}$\pm$0.3 &\cellcolor{lightgray}\textbf{50.9}$\pm$0.1 &\cellcolor{lightgray}\textbf{66.6}$\pm$0.1\\
\midrule
\multirow{2}{*}{\rotatebox{90}{ }}&$\Delta$ Int &+0.3&+0.4&+0.1&+0.3&+2.5&1.0&+1.0\\
&$\Delta$ GNN &+0.2&+0.3&+0.2&+0.2&+1.2&+0.7&+0.6\\
\bottomrule
\end{tabular}
}
\end{table}
\subsection{Overall Performance (RQ1)}
The overall performance of Cassette and all baseline models is evaluated on the COLIEE2022, COLIEE2023 and LeCaRDv2 datasets, as presented in Table~\ref{tab:cassette_overall_2022}, Table~\ref{tab:cassette_overall_2023} and Table~\ref{tab:overall_lecardv2}. It can be observed that \textbf{\textit{the highly efficient Cassette consistently outperforms or matches the heavyweight SOTA model CaseLink, while substantially surpassing both traditional two-tower retrieval baselines and advanced embedding large language models in both one-stage and two-stage settings}}.
 For COLIEE2022 and LeCaRDv2, Cassette can match or outperform the teacher GNN CaseLink. While for COLIEE2023, Cassette can successfully distil to improve closely to the teacher retriever with a gap partially due to the dataset difficulty. It is common that in most GNN-to-MLP distillation for node classification or link prediction~\cite{glnn,llp}, the performance under inductive setting is commonly unstable given that the test data may have distribution shift in the structure. LCR task is a typical inductive setting that the query case is unseen during model development. Furthermore, the importance of the structural information in modelling may vary for different datasets, which also leads to the GNN-to-MLP distillation perform in an unstable way under inductive setting.  Nevertheless, the ranking distillation improve the base model performance to outperform or achieve a comparable performance to the heavy yet powerful teacher retriever in both one- and two-stage results across all datasets, validating the effectiveness of Cassette. Compared with other baselines, Cassette can outperform them in a large margin. BM25, as the statistical method, can achieve a relatively high performance compared with neural network models, verifying the informativeness of term frequency. 

For general pretrained language models, such as LEGAL-BERT, Lawformer and MonoT5, the performance cannot match BM25, which indicates that these legal specific datasets are challenging. While BM25 remains highly efficient and effective for fact heavy queries with strong lexical overlap, Cassette demonstrates clear advantages in scenarios that require deeper semantic understanding and legal reasoning. In particular, it excels at distinguishing cases with similar factual narratives but different charges, legal elements, or judicial logic. From a ranking perspective, BM25 primarily relies on surface level term frequency signals, which may lead to suboptimal ordering when lexical similarity does not faithfully reflect legal relevance. In contrast, Cassette leverages semantically enriched representations and structural distillation signals to produce more legally coherent ranking lists, promoting cases that are aligned in legal reasoning rather than merely sharing keywords. By combining offline candidate encoding with lightweight online inference, Cassette achieves near BM25 level query latency while delivering substantially higher retrieval accuracy and ranking quality in complex scenarios. This analysis highlights the irreplaceability of Cassette in large scale legal retrieval systems that demand both efficiency and semantic precision. The fine-tuned language model with case structure modelling and generation, SAILER and PromptCase, can largely improve the two-tower-based language model retrieval results. E5-Mistral-7b-Instruct and Inf-Retriever-V1 are two top-ranked open-source embedding models on the MTEB~\cite{mteb} benchmark for legal domain tasks, outperforming two-tower baselines due to their advanced architectural design and substantially larger model size. Moreover, Qwen3-Embedding-8B, a leading model for general text embedding task, attains performance comparable to E5-Mistral-7B-Instruct and Inf-Retriever-V1, owing to its extensive pretraining on large and diverse corpora that confer strong generalization capabilities across specialized domains such as law. CaseGNN utilises GNN to encode individual cases, still falling into the two-tower retrieval method category, performs the best in this manner. CaseGNN is not available in Chinese so that it is not evaluated on LeCaRDv2. While CaseLink, serving as the teacher retriever model in this paper, takes advantage from the case-to-case relationships in the case graph to achieve a large improvement in case retrieval. But the retrieval efficiency of CaseLink is unreasonable for real-world deployment. Specifically, $\Delta$ Init represents the performance gap between the feature initialization model (CaseGNN model for COLIEE2022 and COLIEE2023, E5-Mistral-7b-Instruct model for LeCaRDv2) and the proposed Cassette framework. The substantial $\Delta$ Init observed across all three datasets demonstrates that Cassette significantly enhances retrieval effectiveness over the initial embedding features by effectively distilling structural and ranking knowledge from the teacher model. In contrast, $\Delta$ GNN denotes the performance gap between the state-of-the-art GNN-based model, CaseLink, and Cassette. This gap illustrates that Cassette can achieve comparable or even superior performance while offering significantly improved efficiency. These results confirm that Cassette strikes a strong balance between accuracy and scalability, making it well-suited for real-world legal case retrieval scenarios. \textbf{\textit{Cassette can outperform both the existing traditional two-tower retrieval baselines and advanced embedding large language models and achieve higher or comparable results as CaseLink with a much higher efficiency}}.

\subsection{Inference Efficiency (RQ2)}
This experiment will evaluate the retrieval efficiency, which is one of the most important factors in real-world LCR system. The comparison will be evaluated on the inference time when a new query arrives. Cassette is compared with (1) \textbf{Init-feat}, which comes from the encoding results of the best two-tower baseline CaseGNN and is the feature initialisation for Cassette and CaseLink; (2) \textbf{CaseLink}, the teacher retriever model, which is based on heavy online computation with case graph construction and GNN message passing.

According to the inference efficiency results in Table~\ref{tab:efficiency} and Figure~\ref{fig:motivation}, it is demonstrated that \textbf{\textit{Cassette can achieve a 50 times speed-up on the COLIEE dataset and a $\sim$340,000 times speed-up on the LeCaRDv2 dataset compared with the CaseLink while outperforms or achieves a comparable result}}. The speed of Cassette is similar to the traditional two-tower method, while Cassette achieves a much higher performance. The slow speed of CaseLink is due to its $O(n^2)$ complexity and GNN neighbourhood aggregation. It is evident that Cassette achieves significantly greater speed-up as the number of candidate cases increases from 1,563 in COLIEE2022 to 55,192 in LeCaRDv2. The offline calculation for candidate cases and the lightweight student query encoder of online calculation for query cases are the main reasons in the high efficiency of Cassette.

\begin{table}[!t]\centering
\caption{Inference time comparison against retrieval performance. A more intuitive comparison is in Figure~\ref{fig:motivation}.}\label{tab:efficiency}
\small
\resizebox{0.7\linewidth}{!}{
    \begin{tabular}{c|c|c|c|c}
    \toprule
    Dataset&Metric&Init-Feat (CaseGNN)&CaseLink&Cassette\\
    \midrule
    \multirow{2}{*}{COLIEE2022}&Time (s)&0.010 (1$\times$)& 0.529 (\textbf{51.4$\times$})&0.011 (1.1$\times$)\\
    &Ma-F1&42.4&44.2&44.6\\
    \midrule
    \multirow{2}{*}{LeCaRDv2}&Time (s)&0.010 (1$\times$)& 3,572.412 (\textbf{3.6$\text{e}^5\times$})&0.011 (1.1$\times$)\\
    &Ma-F1 &19.4 &20.7 &21.1\\
    \bottomrule
    \end{tabular}
    }
\end{table}

\subsection{Ablation Study (RQ3)}
In ablation study, the effectiveness of different modules will be evaluated. There are three key component modules in Cassette, the student candidate encoder, the student query encoder, and the ranking distillation objective. The following variants are evaluated in this experiment correspondingly: (1) \textbf{Teacher} is the pretrained teacher retriever; (2) \textbf{GNN (ideal)} distils a same-structure GNN as the teacher retriever with Teacher Legal Graph and structural distillation, which is the ideal distillation for retrieval performance with low efficiency issue; (3) \textbf{MLP} uses a same MLP model for both the query and candidate cases for inference without GNN with structural distillation; (4) \textbf{Sup} only uses the supervised loss in Eq.~\eqref{eq:infonce} to train the student candidate encoder and query encoder without any distillation. (5) \textbf{Sup+Ranking} leverages both supervised loss and ranking distillation loss in Eq.~\eqref{eq:ranking_match} to train both the student query and candidate encoders. (6) \textbf{Sup+Eigen} employs supervised loss and eigen-matching distillation loss in Eq.~\eqref{eq:eig_match} to train both the student query and candidate encoders.

\begin{table*}[!t]\centering
\caption{Ablation study on COLIEE datasets. (\%)}\label{tab:ablation}
\resizebox{0.9\linewidth}{!}{
\begin{tabular}{l|ccccccc}
\toprule
\midrule
\multirow{2}{*}{Variants} &\multicolumn{7}{c}{COLIEE2022} \\
\cmidrule{2-8}
&P@5 &R@5 &Mi-F1 &Ma-F1 &MRR@5 &MAP &NDCG@5 \\
\midrule\midrule
Teacher &37.0$\pm$0.1&43.9$\pm$0.1&40.1 $\pm$0.1&44.2$\pm$0.1&67.3 $\pm$0.5&65.0 $\pm$0.2&70.3 $\pm$0.1\\
GNN (ideal)&36.2$\pm$0.2&43.0$\pm$0.3&39.3$\pm$0.2&43.4$\pm$0.1&65.1$\pm$0.7&62.5$\pm$0.5&68.1$\pm$0.4\\
\midrule
MLP &35.3$\pm$0.2&42.0$\pm$0.2&38.4$\pm$0.2&42.6$\pm$0.2&63.6$\pm$0.4&61.3$\pm$0.3&66.9$\pm$0.3\\
Sup &36.9$\pm$0.1&43.8$\pm$0.1&40.0$\pm$0.1&44.4$\pm$0.1&65.3$\pm$0.1&63.2$\pm$0.2&68.7$\pm$0.1\\
Sup+Ranking &37.1$\pm$0.1&44.0$\pm$0.1&40.3$\pm$0.1&44.5$\pm$0.1&66.6$\pm$0.3&64.0$\pm$0.2&69.5$\pm$0.2\\
Sup+Eigen &37.1$\pm$0.1&44.1$\pm$0.1&40.3$\pm$0.1&44.6$\pm$0.1&66.0$\pm$0.2&63.7$\pm$0.1&69.0$\pm$0.2\\
\rowcolor{lightgray}
Cassette &37.2$\pm$0.1 &44.2$\pm$0.2 &40.4$\pm$0.1 &44.6$\pm$0.1 &66.9$\pm$0.1 &64.4$\pm$0.1 &69.7$\pm$0.1\\
\bottomrule
\multicolumn{8}{c}{ }\\
\toprule
\midrule
\multirow{2}{*}{Variants} &\multicolumn{7}{c}{COLIEE2023} \\
\cmidrule{2-8}
&P@5 &R@5 &Mi-F1 &Ma-F1 &MRR@5 &MAP &NDCG@5 \\\midrule
\midrule
Teacher &20.9 $\pm$0.3&38.4 $\pm$0.6&27.1$\pm$0.3&28.2 $\pm$0.3&45.8 $\pm$0.5&44.3 $\pm$0.7&49.8 $\pm$0.4\\
GNN (ideal) &23.1$\pm$0.4&38.9$\pm$0.3&27.6$\pm$0.8&28.1$\pm$0.3&44.1$\pm$0.6&42.6$\pm$0.4&47.9$\pm$0.4\\
\midrule
MLP &16.9$\pm$0.3&31.4$\pm$0.5&22.0$\pm$0.4&22.7$\pm$0.4&36.8$\pm$0.2&35.7$\pm$0.2&40.9$\pm$0.3\\
Sup &19.5$\pm$0.1&36.3$\pm$0.1&25.4$\pm$0.1&26.3$\pm$0.1&41.2$\pm$0.3&39.8$\pm$0.4&45.6$\pm$0.2\\
Sup+Ranking &19.7$\pm$0.1&36.5$\pm$0.2&25.6$\pm$0.2&26.4$\pm$0.2&41.3$\pm$0.3&40.0$\pm$0.5&45.8$\pm$0.2\\
Sup+Eigen &19.7$\pm$0.1&36.5$\pm$0.2&25.6$\pm$0.1&26.4$\pm$0.1&41.7$\pm$0.8&40.5$\pm$0.7&46.1$\pm$0.6\\
\rowcolor{lightgray}
Cassette &19.8$\pm$0.1 &36.7$\pm$0.2 &25.7$\pm$0.1 &26.6$\pm$0.2 &42.5$\pm$0.2 &41.1$\pm$0.1 &46.8$\pm$0.2 \\
\bottomrule
\end{tabular}}
\end{table*}

\vspace{0.5cm}
\begin{table}[!t]\centering
\caption{Ablation study on LeCaRDv2 dataset. (\%)}\label{tab:ablation_lecardv2}
\resizebox{0.9\linewidth}{!}{
\begin{tabular}{l|ccccccc}
\toprule\midrule
\multirow{2}{*}{Variants} &\multicolumn{7}{c}{LeCaRDv2}\\
\cmidrule{2-8}
&P@30 &R@30 &Mi-F1 &Ma-F1 &MRR@30 &MAP &NDCG@30 \\
\midrule\midrule
Teacher &19.1$\pm$0.1 &23.3$\pm$0.1 &21.0$\pm$0.3 &20.7$\pm$0.1 &63.3$\pm$0.3 &44.0$\pm$0.2 &62.3$\pm$0.4\\
GNN (ideal) &19.2$\pm$0.1 &23.5$\pm$0.1 &21.1$\pm$0.1 &20.8$\pm$0.1 &63.2$\pm$0.1 &44.1$\pm$0.2 &62.4$\pm$0.1 \\
\midrule
MLP &17.7$\pm$0.1&21.5$\pm$0.1&19.5$\pm$0.2&19.1$\pm$0.1&58.4$\pm$0.3&40.3$\pm$0.2&59.1$\pm$0.1\\
Sup &17.0$\pm$0.2&20.8$\pm$0.2&18.7$\pm$0.2&18.5$\pm$0.3&56.9$\pm$0.1&40.1$\pm$0.9&58.3$\pm$0.8\\
Sup+Ranking &17.2$\pm$0.3&21.1$\pm$0.3&18.9$\pm$0.3&18.6$\pm$0.3&57.3$\pm$0.6&40.4$\pm$0.8&59.1$\pm$0.8\\
Sup+Eigen &17.6$\pm$0.6&21.5$\pm$0.8&19.3$\pm$0.7&19.0$\pm$0.7&58.0$\pm$0.6&40.1$\pm$0.4&58.4$\pm$0.9\\
\rowcolor{lightgray}
Cassette &19.5$\pm$0.1 &23.9$\pm$0.1 &21.4$\pm$0.1 &21.1$\pm$0.1 &63.7$\pm$0.9 &45.0$\pm$0.4 &63.4$\pm$0.3\\
\bottomrule
\end{tabular}}
\end{table}

According to Table~\ref{tab:ablation} and~\ref{tab:ablation_lecardv2}, it can be seen that \textbf{\textit{the student encoders, as well as the ranking distillation in Cassette have a significant contribution}} to the retrieval performance with the knowledge transfer from the teacher retriever to Cassette. The Teacher  variant serves as the  teacher model for knowledge transfer. The GNN (ideal) variant provides an ideal distillation performance by using a same structure GNN to distil the knowledge from the teacher retriever. By only distilling one student encoder for both candidate and query cases, the performance of the student encoder is not improved even after structural distillation, which indicates that the structure information among cases cannot be directly distilled into the structure agnostic MLP models. By removing the structural distillation objective and only using the supervised loss to train the two student encoders, the Sup variant cannot match the performance of teacher retriever and Cassette in most of the metrics in three datasets, which verifies that the proposed structural distillation objective can transfer the knowledge from the teacher retriever and trains the two student encoders more effectively than the supervised objective. On the the three datasets, only the use of ranking or eigen-matching distillation surpasses the performance of the MLP baseline in most of the metrics, highlighting the importance of structural information and the effectiveness of distillation techniques, which are absent in the MLP model.

\begin{figure}[!t]
\centering
    \subfigure{
    \includegraphics[width=0.47\linewidth]{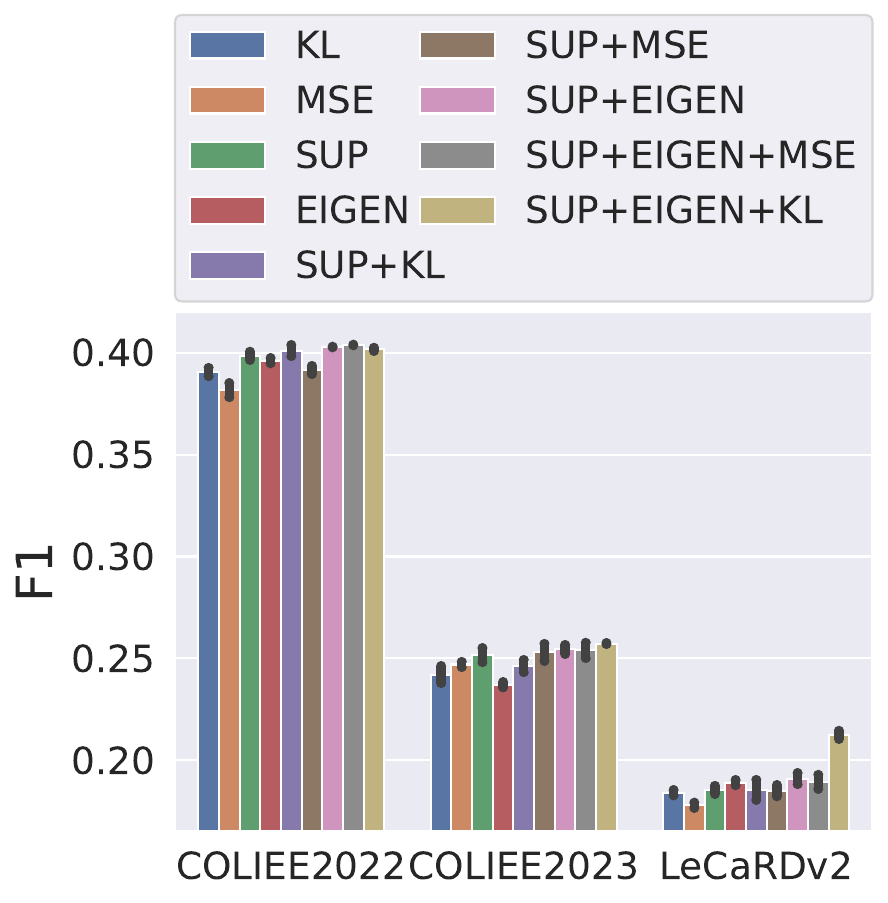}
    \label{fig:dis-obj-f1}
    }
    \subfigure{
    \includegraphics[width=0.47\linewidth]{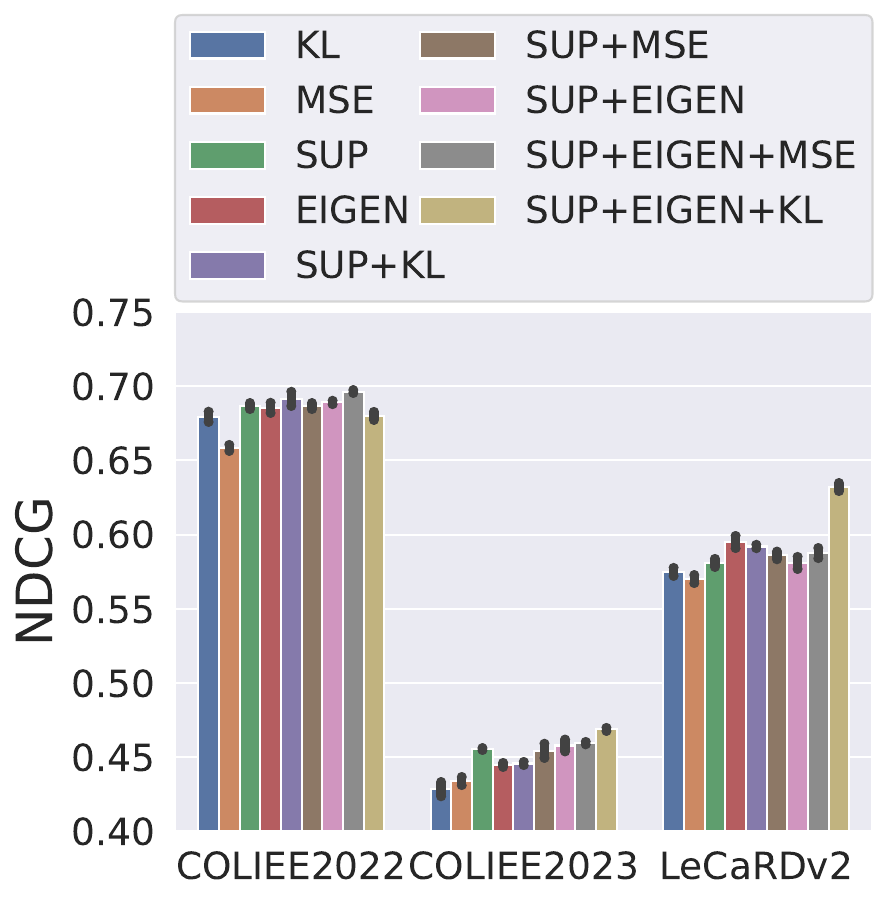}
    \label{fig:dis-obj-ndcg}
    }
    \vspace{-0.3cm}
\caption{Different distillation objectives.}
\label{fig:dis-obj}
\end{figure}

\subsection{Effectiveness of Structural Distillation Objective (RQ4)}
This experiment evaluates the effectiveness of different structural distillation objectives. The ranking objective in Eq.~\eqref{eq:ranking_match} can be implemented in (1) \textbf{KL} divergence between the Softmax-normalised ranking scoresand or (2) \textbf{MSE} between the ranking scores; (3) \textbf{SUP} as the supervised loss in Eq.~\eqref{eq:loss-overall} using ground-truth label, the same as -Sup in Ablation Study; (4) \textbf{EIGEN} is the eigen-matching objective; (5) \textbf{SUP+KL}; (6) \textbf{SUP+MSE}; (7) \textbf{SUP+EIGEN+MSE}; (8) \textbf{SUP+EIGEN+KL}.

As in Figure~\ref{fig:dis-obj}, either combining the supervised loss and the structural distillation or directly using the structural distillation yield improved results, indicating that \textbf{\textit{the soft target from the teacher retriever provides valuable information for training student models}}. Although KL divergence is widely used for knowledge distillation, the performance is inferior to other objectives. It is observed that eigen-matching is not so useful on its own, but the performance improves when supervised loss and ranking loss are incorporated, which shows that combining structural alignment with task-specific supervision enhances the model’s ability to capture meaningful case relationships.

\subsection{Comparison between Student Candidate Case Encoder and Teacher Retriever (RQ5)}
\label{sec:comparisonteachernstudentgnn}

This experiment investigates the performance differences between using the proposed student candidate encoder and the teacher GNN retriever for encoding candidate cases during testing on the Student Candidate Legal Graph (SCLG). The goal is to understand the behaviour and generalisation capability of graph-based models under varying test time graph structures. In Table~\ref{tab:comparison}, \textbf{Teacher GNN} refers to the setting where the teacher GNN is used as the candidate case encoder during test time inference, while \textbf{Cassette} refers to the setting where the proposed student candidate encoder is employed.

As shown in Table~\ref{tab:comparison} and Table~\ref{tab:comparison-lecard}, \textbf{\textit{applying the teacher retriever during testing results in a performance drop}}. This decline occurs because the teacher model is trained on a graph that includes both candidate and query cases, whereas the test time graph (SCLG) contains only candidate cases. The mismatch in graph structure between training and testing phases undermines the encoding effectiveness of the teacher GNN when directly applied during inference.

The teacher GNN directly encodes explicit graph structures, which may include both meaningful relational signals and structural noise such as spurious or degree-driven connections. During distillation, the student learns from softened relational supervision rather than raw graph propagation. This process acts as a selective information filter where salient structural patterns are retained, while noisy or low-confidence signals are attenuated. The consistent gains in ranking metrics like MRR, MAP and NDCG suggest that this filtering improves fine-grained ordering consistency and generalization.

The improvements are stable across datasets with different difficulty levels and relevance distributions. Gains are more pronounced on COLIEE2023 and LeCaRDv2, where the distribution is more challenging or the candidate space is larger. This indicates that the distilled student is less sensitive to distribution shifts and long-tail retrieval settings. The cross-dataset consistency strengthens the robustness and reproducibility of the method.

The teacher GNN is tightly coupled with graph propagation, which may introduce over-smoothing or structural bias. In contrast, the student architecture retains distilled relational inductive biases while being structurally simpler and more adaptable to retrieval objectives. This combination of compressed structural knowledge and flexible optimization likely contributes to the systematic improvements across both classification and ranking metrics.

It is worth noting that if the Teacher Legal Graph contains both query and candidate nodes and is dynamically reconstructed for each test query, the teacher retriever can maintain strong performance. However, this assumption incurs significant computational overhead, as demonstrated by the CaseLink model, and is impractical for realistic applications due to the repeated graph construction and message passing required at test time. These results highlight the practical advantages of Cassette’s student candidate encoder, which maintains competitive accuracy while offering significantly better scalability and generalization under realistic test time constraints.
\begin{table*}[!t]\centering
\caption{Comparison between testing SCLG encoding with Cassette and teacher GNN for COLIEE. (\%)}\label{tab:comparison}
\vspace{-0.3cm}
\resizebox{0.8\linewidth}{!}{
\begin{tabular}{l|ccccccc}
\toprule\midrule
\multirow{2}{*}{Methods} &\multicolumn{7}{c}{COLIEE2022}\\
\cmidrule{2-8}
&P@5 &R@5 &Mi-F1 &Ma-F1 &MRR@5 &MAP &NDCG@5 \\
\midrule\midrule
Teacher GNN &36.1 &42.9 &39.2 &43.0 &65.4 &62.9 &67.9 \\
\midrule
Cassette &37.2 &44.2 &40.4 &44.6 &66.9 &64.4 &69.7 \\
\bottomrule
\multicolumn{8}{c}{ }\\
\toprule
\midrule
\multirow{2}{*}{Methods} &\multicolumn{7}{c}{COLIEE2023} \\
\cmidrule{2-8}
&P@5 &R@5 &Mi-F1 &Ma-F1 &MRR@5 &MAP &NDCG@5 \\\midrule
\midrule
Teacher GNN &17.6 &32.7 &22.9 &23.8 &37.7 &36.7 &42.0\\
\midrule
Cassette &19.8 &36.7 &25.7 &26.6 &42.5 &41.1 &46.8\\
\bottomrule
\end{tabular}}

\end{table*}
\begin{table}[!t]\centering
\caption{Comparison between testing SCLG encoding with Cassette and teacher GNN for LeCaRDv2. (\%)}\label{tab:comparison-lecard}
\resizebox{0.8\linewidth}{!}{
\begin{tabular}{l|ccccccc}
\toprule\midrule
\multirow{2}{*}{Methods} &\multicolumn{7}{c}{LeCaRDv2}\\
\cmidrule{2-8}
&P@30 &R@30 &Mi-F1 &Ma-F1 &MRR@30 &MAP &NDCG@30 \\
\midrule\midrule
Teacher GNN &16.9 &20.7 &18.6 &18.6 &56.8 &41.5 &59.4\\
\midrule
Cassette &19.5&23.9&21.4&21.1&63.7&45.0&63.4\\
\bottomrule
\end{tabular}}
\end{table}

\subsection{Effectiveness of Incremental Document Insertion without Model Retraining (RQ6)}

\begingroup
\begin{table}[t]
\centering
\caption{Retrieval performance before and after document insertion on COLIEE2023.}
\label{tab:document_insertion}
\begin{tabular}{l c c c c c c c}
\toprule
Dataset & P@5 & R@5 & Mi-F1 & Ma-F1 & MRR@5 & MAP & NDCG@5  \\
\midrule
COLIEE2023 &19.8 &36.7 &25.7 &26.6 &42.5 &41.1 &46.8  \\
Mixed-COLIEE2023 & 18.8 & 34.8 & 24.4 & 25.2 & 40.2 & 38.9 & 44.2 \\
\bottomrule
\end{tabular}
\end{table}
\endgroup

To study the effect of document insertion on retrieval performance, we conducted an experiment that simulates incremental corpus growth without retraining the retrieval model. Since both training and test sets of our datasets are all in a fixed setting, with no document insertion during training or testing, a \textbf{Mixed-COLIEE2023} setting is designed by merging the COLIEE2023 training cases into the original test candidate pool. The same pretrained Cassette model and retrieval configuration are applied, and no additional fine-tuning is performed. This setup therefore reflects a realistic document insertion scenario in which new documents are inserted into the retrieval corpus while the underlying model remains fixed.

Table~\ref{tab:document_insertion} reports the retrieval results before and after document insertion. Compared to the original COLIEE2023 setting, retrieval performance under the mixed corpus exhibits a consistent but moderate decline across all metrics. This degradation is expected due to the enlarged and more competitive candidate pool, and it indicates increased retrieval difficulty rather than model instability. These results suggest that the proposed framework is robust to document insertion, maintaining competitive retrieval accuracy even as the corpus size increases. 

\subsection{Robustness Against Charge Category (RQ7)} 
\begingroup
\begin{table}[t]
\centering
\caption{NDCG@5 Performance (\%) by Legal Charge Group on COLIEE2022.}
\label{tab:ndcg_by_group_2022}
\begin{tabular}{lccc}
\hline
\textbf{Group} & \textbf{\# Query} & \textbf{CaseLink} & \textbf{Cassette} \\
\hline
Criminal\_Security\_Public\_Safety &93 & 57.8 & 59.2 \\
Courts\_Procedure\_Administrative\_Law &203 & 72.8 & 73.0 \\
Immigration\_Citizenship\_Border &3 & 66.7 & 64.9 \\
Information\_IP\_Communications\_Other &1 & 1.0 & 1.0 \\
\hline
\end{tabular}
\end{table}
\endgroup

\begingroup
\begin{table}[t]
\centering
\caption{NDCG@5 Performance (\%) by Legal Charge Group on COLIEE2023.}
\label{tab:ndcg_by_group_2023}
\begin{tabular}{lccc}
\hline
\textbf{Group} &\textbf{\# Query}& \textbf{CaseLink} & \textbf{Cassette} \\
\hline
Criminal\_Security\_Public\_Safety &87 & 50.1 & 44.2 \\
Courts\_Procedure\_Administrative\_Law &220 & 47.2 & 45.2 \\
Immigration\_Citizenship\_Border &5 & 51.4 & 31.5 \\
Finance\_Banking\_Tax\_Pensions &1 & 0.0 & 0.0 \\
Agriculture\_Fisheries\_Food &1 & 43.1 & 0.0 \\
Information\_IP\_Communications\_Other &5 & 41.2 & 52.6 \\
\hline
\end{tabular}
\end{table}
\endgroup

To examine whether the proposed framework exhibits bias toward specific legal domains, robustness is evaluated across ground-truth charge categories. Tables~\ref{tab:ndcg_by_group_2022} and \ref{tab:ndcg_by_group_2023} report NDCG@5 performance grouped by legal charge types on COLIEE2022 and COLIEE2023.

Across both datasets, performance remains stable in the major categories with substantial query counts. In Criminal\_Security\_Public\_Safety and Courts\_Procedure\_Administrative\_Law, which together account for the majority of queries, CaseLink achieves results that are either comparable to or consistently better than the baseline. Notably, on COLIEE2023, clear improvements are observed in both major groups, indicating stronger generalization under more challenging settings. For Immigration\_Citizenship\_Border, although the number of queries is relatively small in both years, performance remains competitive and even shows a substantial advantage on COLIEE2023. Categories with only one or a few queries (e.g., Finance\_Banking\_Tax\_Pensions and Agriculture\_Fisheries\_Food) display higher variance, which is expected due to limited statistical support.

Overall, the results demonstrate that performance gains are not confined to a particular charge category. Instead, the framework maintains balanced effectiveness across diverse legal domains, suggesting strong robustness and reduced sensitivity to case-type distribution shifts.

\subsection{Hyperparameter Sensitivity (RQ8)}
This experiment evaluates the sensitivity of Cassette to the weight coefficients for ranking matching and eigen-matching objectives, denoted as $\alpha$ and $\beta$ in Eq.~\eqref{eq:loss-overall}.

\begin{figure}[!t]
\centering
    \subfigure{
    \includegraphics[width=0.47\linewidth]{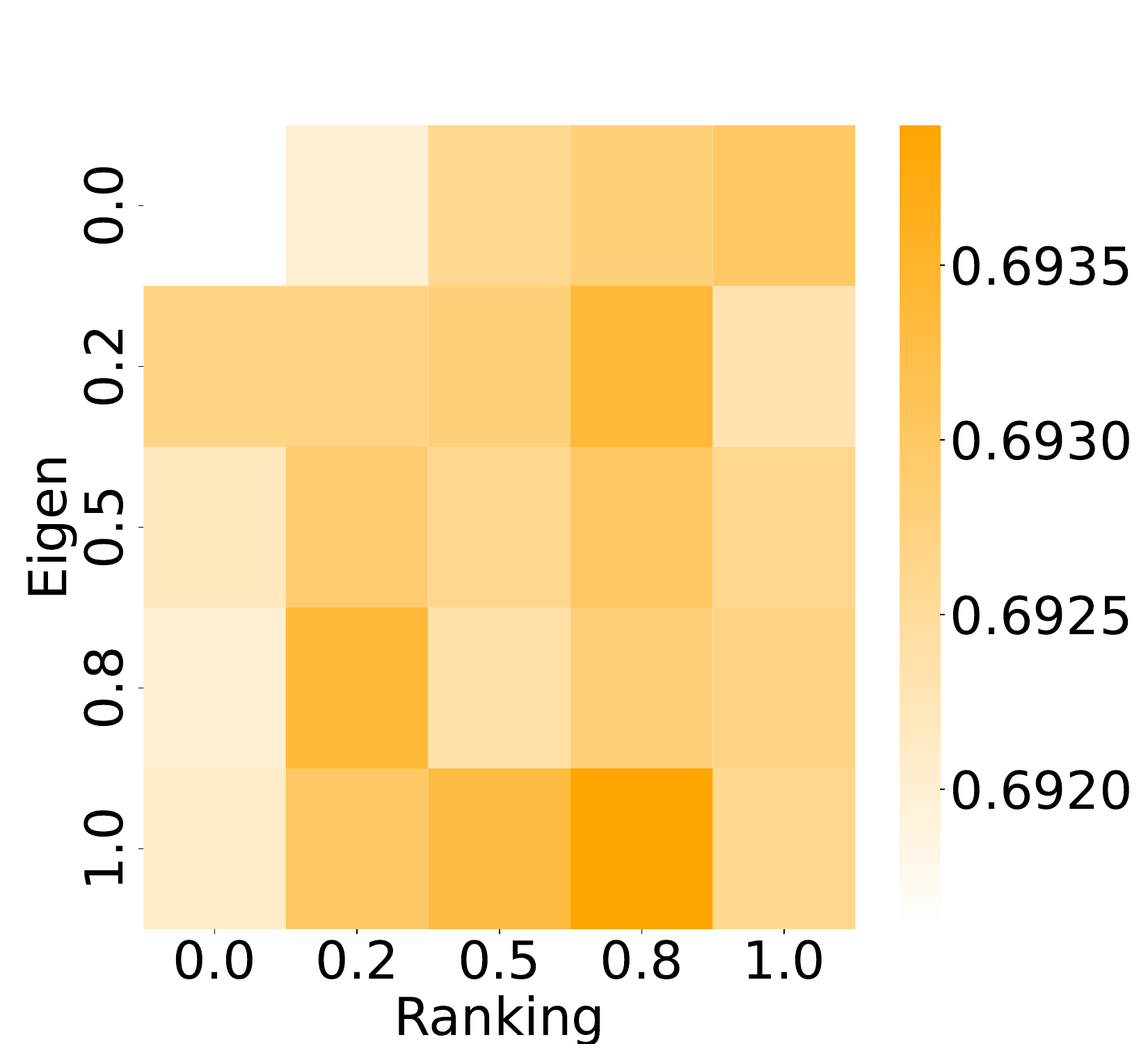}
    \label{fig:lambda-f1}
    }
    \subfigure{
    \includegraphics[width=0.47\linewidth]{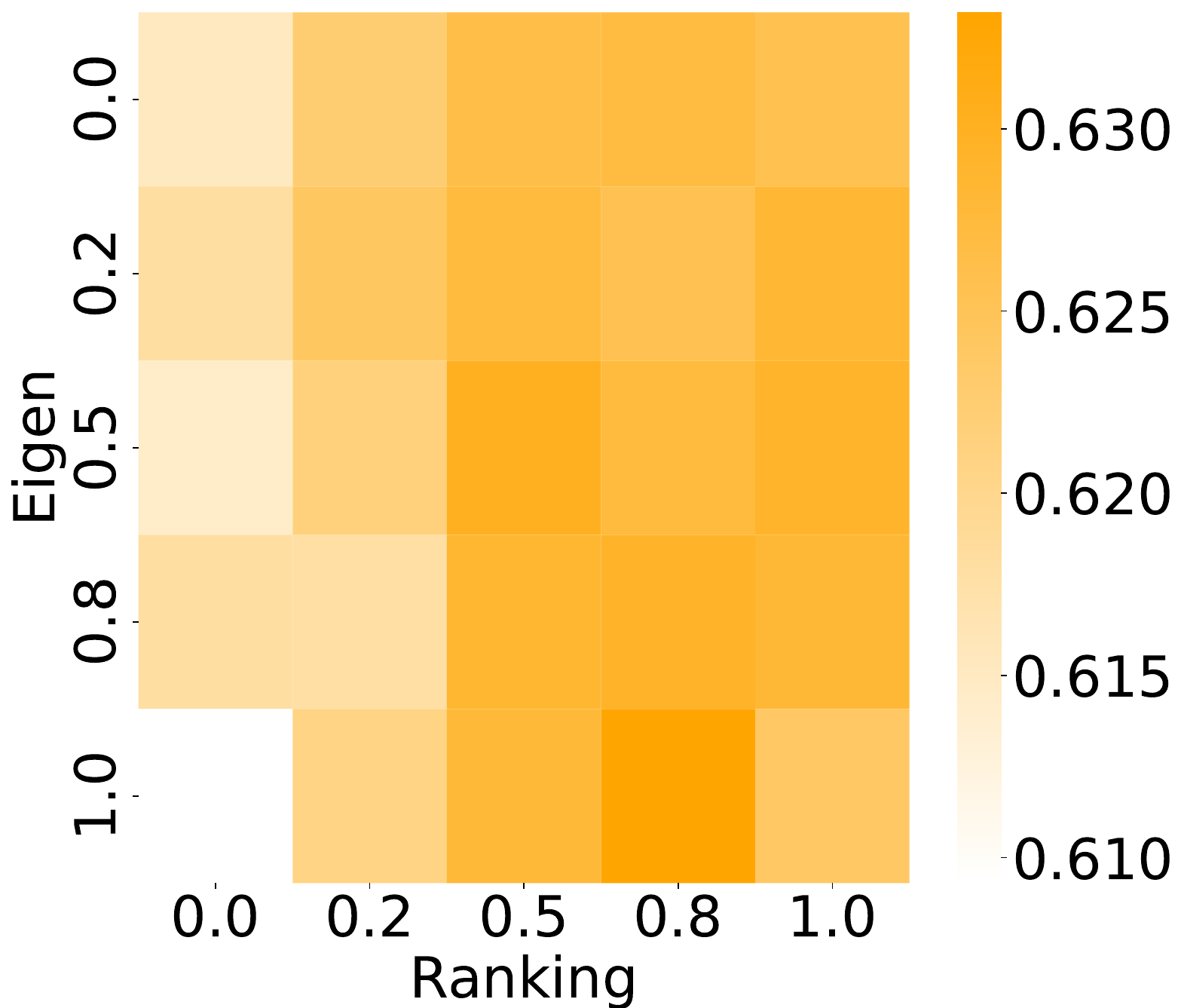}
    \label{fig:lambda-ndcg}
    }
\caption{Results with different weight scales for the ranking matching and the eigen-matching on COLIEE2022 (left) and LeCaRDv2 (right). A reasonable combination of distillation objectives can consistently improve the results.}
\label{fig:lambda}
\end{figure}

As shown in Figure~\ref{fig:lambda}, Cassette demonstrates consistent improvements when both ranking and eigen-matching objectives are assigned non-trivial weights during training. On both COLIEE2022 and LeCaRDv2 datasets, the retrieval performance peaks when $\alpha$ and $\beta$ are balanced, indicating that the synergy between ranking-based supervision and structural alignment is essential for effective knowledge transfer.

In contrast, when either objective is omitted (i.e., setting $\alpha = 0$ or $\beta = 0$), there is a noticeable decline in performance. This suggests that relying solely on one type of supervision will lead to suboptimal learning of the student encoders. These results validate the importance of incorporating both ranking semantics and structural consistency in the distillation process. Thus, the sensitivity analysis confirms that Cassette is robust across a reasonable range of $\alpha$ and $\beta$ values, and its effectiveness stems from the complementary roles of the ranking and structural objectives in capturing the complex dependencies inherent in legal case retrieval.

In Figure~\ref{fig:lambda}, $\mu$ is fixed to 1 to isolate the interaction between $\alpha$ and $\beta$. Under $\mu = 1$, the distillation signal provides stable and balanced guidance, enabling us to examine the structural–ranking trade-off without introducing an additional scaling factor. The performance surface over $\alpha$ and $\beta$ exhibits a smooth, plateau-like region rather than sharp isolated peaks, indicating that the objective function is well-conditioned around its optimum. This behavior suggests that $\alpha$ primarily regulates structural smoothness and global consistency, while $\beta$ governs discriminative ranking alignment. The structural regularization term and the ranking-aware loss therefore impose complementary constraints on representation learning: the former enforces global coherence and mitigates variance through spectral smoothing, whereas the latter enhances task-specific discrimination and prevents excessive smoothing. The empirically optimal region emerges in a balanced regime where structural priors shape representation geometry without overwhelming discriminative learning signals.

Moreover, the slight shift of optimal combinations across datasets reflects differences in graph density and relational reliability. When structural correlations are strong, a relatively larger $\alpha$ is beneficial; when ranking supervision is more informative, increasing $\beta$ becomes advantageous. Similarly, $\mu$ controls the strength of distillation regularization and typically achieves the best performance within an intermediate range, where knowledge transfer stabilizes training without restricting adaptive capacity. Overall, the smooth performance landscape and broad near-optimal region demonstrate that the framework is theoretically grounded in a regularization–discrimination trade-off and remains robust under moderate hyperparameter variations.

\section{Conclusion}
This paper tackles the efficiency limitations of our previous work, CaseLink~\cite{caselink}, a graph-based legal case retrieval (LCR) methods, particularly those arising from dynamic graph construction and computationally intensive GNN operations during inference. To overcome these challenges, this paper proposes an extension based on novel distillation framework, Cassette, which balances retrieval effectiveness with high efficiency and practical scalability.

Cassette introduces a hybrid student dual encoder architecture trained under the supervision of a powerful graph-based teacher retriever. Specifically, to eliminate the need for repeated graph reconstructions when handling new queries, a \textit{student candidate case encoder} is developed to process all candidate cases offline. In parallel, a lightweight \textit{student query encoder} is designed for efficient real-time inference. The student encoders are optimized using a combination of ranking and eigen-matching objectives, ensuring that both relevance and structural information are effectively transferred from the teacher model.

Extensive multi-lingual experiments on COLIEE2022, COLIEE2023, and LeCaRDv2 demonstrate that Cassette not only achieves retrieval performance comparable to the state-of-the-art, but also offers substantial improvements in inference efficiency, making it suitable for real world deployment in large scale legal information systems.

\section{Future Work}
Future research can extend Cassette in several directions. First, adapting the framework to support multilingual or cross-jurisdictional legal corpora would enhance its applicability in global legal systems. Second, integrating domain specific prompts or reasoning chains into the student encoders may further boost interpretability and semantic alignment. Additionally, exploring online or continual distillation strategies could enable Cassette to incrementally adapt to evolving legal databases without retraining from scratch, further improving its scalability and practicality in dynamic legal environments.

Another promising direction for future work is the integration of fine-grained legal knowledge into the current framework. By extending the system with external legal knowledge sources, such as statutes, case precedents, and legal commentaries, the model could capture more nuanced legal reasoning and improve its understanding of complex legal relationships. Incorporating such structured and domain-specific knowledge has the potential to enhance both the accuracy and interpretability of legal document retrieval and analysis, making the framework more robust and practically applicable in real-world legal scenarios. The teacher GNN can then encode this legal structured information, and the student model can inherit this information through structurally guided distillation. This design allows the distillation process to move beyond generic representation transfer and instead reflect core principles of judicial reasoning. For example, provision constraints encode normative boundaries that restrict charge applicability. When incorporated into the teacher’s relational supervision, these signals can guide the student to learn consistency patterns across similar charges, analogous factual scenarios, and legally constrained outcomes.

\begin{acks}
This work is supported by Australian Research Council CE200100025, DP230101196, DP230101753, DE250100919, LP230200892, LP240200546.
\end{acks}

\bibliographystyle{ACM-Reference-Format}
\bibliography{acmart}

\end{document}